# PHIBSS: Unified Scaling Relations of Gas Depletion Time and Molecular Gas Fractions [1]

L.J. Tacconi[1], R. Genzel[1,2,3], A. Saintonge[4], F. Combes[5], S. García-Burillo[6], R. Neri[7], A. Bolatto[8], T. Contini[9], N.M. Förster Schreiber[1], S. Lilly[10], D. Lutz[1], S. Wuyts[11], G. Accurso[4], J. Boissier[7], F. Boone[9], N. Bouché[9], F. Bournaud[12], A. Burkert[13,1], M. Carollo[10], M. Cooper[14], P. Cox[15], C. Feruglio[16], J. Freundlich[5,17], R. Herrera-Camus[1], S. Juneau[12], M. Lippa[1], T. Naab[18], A. Renzini[19], P. Salome[5], A. Sternberg[20], K. Tadaki[1], H. Übler[1], F. Walter[21], B. Weiner[22] & A.Weiss[23]

*[1] Max-Planck-Institut für extraterrestrische Physik (MPE), Giessenbachstr., 85748 Garching, FRG(linda@mpe.mpg.de, genzel@mpe.mpg.de )*
*[2] Dept. of Physics, Le Conte Hall, University of California, 94720 Berkeley, USA*
*[3] Dept. of Astronomy, Campbell Hall, University of California, Berkeley, CA 94720, USA*
*[4]Department of Physics & Astronomy, University College London, Gower Place, London WC1E 6BT, UK*
*[5] Observatoire de Paris, LERMA, College de France, CNRS, PSL Univ., Sorbonne Univ. UPMC, F-75014, Paris, France*
*[6] Observatorio Astronómico Nacional-OAN, Observatorio de Madrid, Alfonso XII, 3, 28014 - Madrid, SP*
*[7]IRAM, 300 Rue de la Piscine, 38406 St.Martin d'Heres, Grenoble, France*
*[8] Dept. of Astronomy, University of Maryland, College Park, MD 20742-2421, USA*
*[9] Institut d'Astrophysique et de Planétologie, Universite de Toulouse, 9 Avenue du Colonel Roche BP 44346 - 31028 Toulouse Cedex 4, FR*
*[10] Institute of Astronomy, Department of Physics, Eidgenössische Technische Hochschule, ETH Zürich, CH-8093, SW*
*[11]Department of Physics, University of Bath, Claverton Down, Bath, BA2 7AY, United Kingdom*
*[12] Service d'Astrophysique, DAPNIA, CEA/Saclay, F-91191 Gif-sur-Yvette Cedex, FR*
*[13] Universitätssternwarte der Ludwig-Maximiliansuniversität, Scheinerstr. 1, D-81679 München, FRG*
*[14]Department of Physics and Astronomy, Frederick Reines Hall, University of California, Irvine, CA 92697, USA*
*[15]ALMA Santiago Central Office, Alonso de Córdova 3107, Vitacura, Santiago, CH*
*[16]Scuola Normale Superiore di Pisa, IT*
*[17] Racah Institute of Physics, The Hebrew University, Jerusalem 91904, Israel*
*[18]Max-Planck-Institut für Astrophysik, Karl Schwarzschildstrasse 1, D-85748 Garching, FRG*
*[19] Osservatorio Astronomico di Padova, Vicolo dell'Osservatorio 5, Padova, I-35122, IT*
*[20] School of Physics and Astronomy, Tel Aviv University, Tel Aviv 69978, Israel*
*[21]Max-Planck-Institut für Astronomie (MPIA), Königstuhl 17, 69117 Heidelberg, FRG*
*[22]Steward Observatory, 933 North Cherry Avenue, University of Arizona, Tucson, AZ 85721-0065, USA*
*[23]Max-Planck-Institut für Radioastronomie (MPIfR), Auf dem Hügel 69, 53121 Bonn, FRG*

---

*[1]Based on observations of an IRAM Legacy Program carried out with the NOEMA , operated by the Institute for Radio Astronomy in the Millimetre Range (IRAM), which is funded by a partnership of INSU/CNRS (France), MPG (Germany) and IGN (Spain).*

## ABSTRACT

This paper provides an update of our previous scaling relations (Genzel et al. 2015) between galaxy integrated molecular gas masses, stellar masses and star formation rates, in the framework of the star formation main-sequence (MS), with the main goal to test for possible systematic effects. For this purpose our new study combines three independent methods of determining molecular gas masses from CO line fluxes, far-infrared dust spectral energy distributions, and ~1mm dust photometry, in a large sample of 1444 star forming galaxies (SFGs) between z=0 and 4. The sample covers the stellar mass range log($M_*$/M$_\odot$)=9.0-11.8, and star formation rates relative to that on the MS, $\delta MS$=$SFR/SFR(MS)$, from $10^{-1.3}$ to $10^{2.2}$. Our most important finding is that all data sets, despite the different techniques and analysis methods used, follow the same scaling trends, once method-to-method zero point offsets are minimized and uncertainties are properly taken into account. The molecular gas depletion time $t_{depl}$, defined as the ratio of molecular gas mass to star formation rate, scales as $(1+z)^{-0.6} \times (\delta MS)^{-0.44}$, and is only weakly dependent on stellar mass. The ratio of molecular-to-stellar mass $\mu_{gas}$ depends on $(1+z)^{2.5}\times (\delta MS)^{0.52}\times (M_*)^{-0.36}$, which tracks the evolution of the specific star formation rate. The redshift dependence of $\mu_{gas}$ requires a curvature term, as may the mass-dependences of $t_{depl}$ and $\mu_{gas}$. We find no or only weak correlations of $t_{depl}$ and $\mu_{gas}$ with optical size R or surface density once one removes the above scalings, but we caution that optical sizes may not be appropriate for the high gas and dust columns at high-z.

## 1. Gas and Galaxy Evolution: Summary of Current State of Research

Throughout the epoch from the peak of the cosmic galaxy/star formation activity ~ 10 Gyrs ago ($z$~2) to the present time the dominant fraction (90%) of the cosmic star formation activity and resulting galaxy growth occurs on a well-defined (dispersion ±0.3 dex), almost linear relation between stellar mass ($M_*$) and star formation rate ($SFR$), the '*star-formation main sequence* ($MS$)' (e.g. Brinchmann et al. 2004, Schiminovich et al. 2007, Noeske et al. 2007, Daddi et al. 2007, Franx et al. 2008, Elbaz et al. 2007, 2011, Rodighiero et al. 2010, 2011, Peng et al. 2010, Whitaker et al. 2012, 2014 (hereafter W12 and W14), Renzini & Peng 2015, Speagle et al. 2014 (hereafter S14), Schreiber et al. 2015).

The resulting dependence of the specific star formation rate on stellar mass, $sSFR=SFR/M_*$, of MS-star forming galaxies ($SFGs$) varies only slowly with stellar mass ($sSFR$~$M_*^{-0.1..-0.4}$), but its zero point increases strongly with redshift, $sSFR \propto (1+z)^3$ to $z$~2, and $\propto (1+z)^{1.5}$ for $z$>2 (Lilly et al. 2013). There are different MS-prescriptions proposed in the literature (e.g. W12, W14, S14), with differences in zero-points and slopes in the mass and redshift relations dependent on sample selections (redshift range, survey bands), survey completeness, and methodologies applied to derive $M_*$ and $SFR$s (Renzini & Peng 2015). Figure 1 shows the MS-lines proposed by S14, W12 and W14 in log$sSFR$-log$(1+z)$ (corrected to a common fiducial stellar mass of 5x10$^{10}$ M$_\odot$) and log$sSFR$-log$M_*$ (corrected to a common redshift of $z$=1.5). As is evident from the left and right panels in this Figure it is particularly important whether star formation rates are inferred from UV plus infrared (24μm, 70-160μm) photometry, which are preferred over $SFR$s derived from SED-synthesis fitting. Unfortunately this excludes all galaxies at z>3 and those with $\delta MS$≤0 (where $\delta MS$ is the offset from the MS line) for $z$>1.5. The 3D HST $SFR$s used in this paper are based on a combination of these methods, called the 'ladder technique' (Wuyts et al. 2011a, section 3.3). We adopt in this study the prescription proposed by S14,

$$\log(sSFR(MS,z,M_*)\ (\mathrm{Gyr}^{-1})) = (-0.16 - 0.026\times t_c(Gyr))\times(\log M_*(M_\odot) + 0.025) \\ -(6.51 - 0.11\times t_c(Gyr))\ +9\quad ,$$

$$\text{with}\quad \log t_c(Gyr) = 1.143 - 1.026\times\log(1+z) - 0.599\times\log^2(1+z) \\ +0.528\times\log^3(1+z) \qquad (1).$$

Here $t_c$ (Gyr) is the cosmic time for a flat Λ CDM Universe with $\Omega_m$=0.3 and $H_0$=70 km/s/Mpc used throughout this paper, and all stellar masses and star formation rates assume a Chabrier (2003) initial stellar mass function. Compared to alternative prescriptions in the literature (e.g. Genzel et al. 2015 (hereafter G15), Renzini and Peng 2015), the S14 prescription has the advantage of being applicable over a wide range of redshift ($z$=0-5) and stellar mass (log($M_*$/M$_\odot$)=9.0-11.8), but it is mainly based on SED-based *SFRs*, which tend to be lower than UV-IR based *SFRs*. Otherwise S14 is similar to the relations proposed by W12 and W14, as seen in Figure 1.

### 1.1. Steady Growth along the Star-Formation Main Sequence

The MS-evolution has been interpreted in terms of an '*equilibrium-growth/gas-regulator model*' (e.g. Bouché et al. 2010, Davé, Finlator & Oppenheimer 2011, 2012, Lilly et al. 2013, Peng & Maiolino 2014, Dekel & Mandelker 2014, Rathaus & Sternberg 2016). High *SFR*s and galaxy growth along the main-sequence are sustained for several Gyr by *a continuous supply of fresh gas from the cosmic web and through mostly minor mergers* maintaining large gas reservoirs for star formation (e.g. Keres et al. 2005, Dekel, Sari & Ceverino 2009). At $z$~1-2.5 MS SFGs double their stellar mass on a typical time scale of ~0.5-1 Gyr, but their growth appears to halt when they reach the Schechter mass, $M_*\sim 10^{10.5\text{-}11}$ M$_\odot$, and they transition to the sequence of passive galaxies, in a process (or processes) termed '(*mass*) *quenching*' (Kauffmann et al. 2003, Conroy & Wechsler 2009, Peng et al. 2010). Simulations suggest that in parallel to the average growth along the MS, SFGs oscillate up and down in *sSFR* across the MS-band (±0.6 dex) on a ~1 Gyr time scale, owing to increased and decreased gas accretion rates, and to internal gas transport and 'compaction' events (Tacchella et al. 2016).

### 1.2. Change in Galaxy and ISM Properties along and across the MS

A number of studies in the past decade indicate that the MS-line demarcates not only the location of the maximum number of galaxies as a function of $\delta MS$ at constant $M_*$ and $z$. It is also the location of disk galaxies in terms of structural ($n_{\mathrm{Sersic}}$~1, e.g. Franx et al. 2008, Wuyts et al. 2011b, Lang et al. 2014) and kinematic ($v_{rot}/\sigma_0$>>1, e.g. Förster Schreiber et al. 2009, Wisnioski et al. 2015) properties, from $z$~0 to 2.5. In contrast, going up from the MS, dust temperatures increase at all redshifts (Elbaz et al. 2011, Nordon et al. 2012, Magnelli et al 2014), the dense gas fraction increases (Gracia-Carpio et al 2011, Lada et al. 2012), and the ratio of FIR cooling line to continuum luminosity drops ("FIR line deficit") indicative of local volumetric changes in ISM properties (Stacey et al. 2010, Gracia-Carpio et al 2011, Herrera-Camus et al. 2018).

The equilibrium growth model predicts a *close connection between specific star formation rates (sSFR), gas fractions, and metallicities as a function of redshift*, with only modest changes as compared to $z$=0 MS-star formation physics (e.g. Elmegreen et al. 2009, Krumholz & Dekel 2010, Elbaz et al. 2010, 2011, Gracia-Carpio et al. 2011, Nordon et al. 2012, Lilly et al. 2013, Peng & Maiolino 2014). In the Milky Way and nearby galaxies most and arguably all star formation occurs in massive ($10^{4\cdots6}$ $M_\odot$), dense ($n(H_2) \sim 10^{2\cdots5}$ $cm^{-3}$) and cold ($T_{gas}$~10–30 K), gravitationally bound 'giant molecular clouds' (GMCs; Solomon et al. 1987, Bolatto et al. 2008, McKee & Ostriker 2007), and not in warm atomic gas (Bigiel et al. 2008, Leroy et al. 2008, Schruba et al. 2011). An important open issue is whether the depletion time for converting molecular gas to stars on galactic scales is set locally within clouds (Krumholz & McKee 2005) or on large galactic scales (Elmegreen 1997, Silk 1997). Another is how gas reservoirs change as a function of redshift, stellar mass, star formation rate, galaxy size/internal structure, gas motions and environmental parameters (e.g. Daddi et al. 2010a, b, Tacconi et al. 2010, 2013, Genzel et al. 2010, 2015, Bouché et al. 2010, Lilly et al. 2013, Davé et al. 2011, 1012, Lagos et al. 2011, 2015a,b, Fu et al. 2012, Popping et al. 2015).

# 2. Scaling Relations for Molecular Gas

The goal of this paper is to synthesize the wealth of data now available in the literature, relate these data to the basic galaxy parameters, and derive the most comprehensive (molecular) gas scaling relations currently available as a function of these parameters. Our approach is to

**1.** select a representative and statistically significant 'benchmark' sub-sample of the overall parent SFG population, covering a wide range in basic galaxy parameters and with a well understood selection function from the parent SFG population, sampled by state of the art panchromatic imaging surveys, and including modest samples of 'outlier', star-bursting galaxies for comparison to increase the range covered in δMS;
2. include and compare all three currently available tracers of molecular mass content;
3. establish fitting functions between the galaxy integrated molecular content and the key global galaxy parameters (stellar mass, star formation rate, redshift).

A completely 'un-biased' survey of gas properties for a determination of the multi-dimensional distribution function in ($z, M_*, \delta MS$)-space is currently not realistic, at least not at high-z, because of the prohibitively long required observing times. Establishing scaling relations to 10% accuracy in the fit parameters still require several thousand galaxies if several independent methods are included. Fortunately, G15 previously established that parameter dependences to first order are not correlated and thus can be separated. As we show in this work, this means scaling relations with 10-20% statistical precision in the fit-parameters can be established with ~1000 galaxies (see also Appendix A1).

## 2.1. A summary of Genzel et al. (2015)

This work builds on the previous analysis by G15; we summarize here the salient points that motivate the present work. G15 (and references therein) took advantage of the finding of many observational studies during the last decade (see sections 1.1 and 1.2) that in the physical framework of the evolution of star forming galaxies along the MS, the most important galaxy integrated, cool & dense ISM properties, namely the *molecular gas content relative to the stellar mass*, $\mu = M_{molgas}/M_*$, and the *molecular gas depletion time*, $t_{depl} = M_{molgas}/SFR$ (Gyr), mainly depend on cosmic time (or redshift), and on the location along and perpendicular to the MS-line at a given redshift. Here $M_{molgas}$ is the total molecular gas mass, including a 36% mass fraction of helium, and a correction for the photo-dissociated surface layers of the molecular clouds that are fully molecular in $H_2$ but dissociated ('dark') in CO (see equation (2), Sternberg

& Dalgarno 1995, Wolfire, Hollenbach & McKee 2010, Bolatto, Wolfire & Leroy 2013). G15 show that the scaling relations for $t_{depl}$ and $\mu_{gas}$ can then be written as *products of functions depending on redshift, stellar mass, and on offset from the MS-line*, $\delta MS = sSFR / sSFR(MS,z,M_*)$), and only indirectly on the absolute value of the *SFR* or *sSFR*. G15 show empirically that this *separation of variables* is justified, since the depletion time to first order does not depend on stellar mass, and the slope $d\log t_{depl}/d\delta MS$ does not depend significantly on *z*. We test and reestablish these fundamental assumptions in section 2.2.2. With the scaling relations for $t_{depl}$ established, molecular gas to stellar mass ratios then follow straightforwardly by multiplying the depletion time with the specific star formation rate of a galaxy, $\mu_{gas} = M_{molgas}/M_* = t_{depl} \times (SFR/M_*)$. Separability is thus possible for the $\mu_{gas}$-dependence as well. G15 also show that good fits are obtained to the scaling relations with a product of power law functions in the variables above, resulting in linear fitting functions in log-log space.

We caution that this empirical conclusion is not unique, and we further explore in this paper, whether more complex fitting functions are required, as the quality of the data improves. We also caution that the parameterization in terms of offset from the MS, which in turn is a function of *z* and $M_*$, is well motivated by physical properties (section 1.2), but is mathematically not unique, in part because MS-recipes vary (Figure 1). G15 explored different MS recipes, and also fitted directly in *logsSFR*, $\log(1+z)$ and $\log(M_*)$ space. The main difference in fitting in *z, sSFR*, $M_*$ from fitting in *z*, $\delta MS$, $M_*$ space is the *interpretation* of the parameter values of the redshift scaling. While the slope $d\log t_{depl}/d\log(1+z)$ at constant $\delta MS$=0 obviously describes the redshift evolution of the entire population, the corresponding slope at constant *sSFR* is a redshift cut at that selected *sSFR* and as such does not have a well-defined meaning (G15). For completeness we include also in this paper fits in *z, sSFR*, $M_*$ space. In terms of $\chi_r^2$ this fit is somewhat worse than that of the $\delta MS$-fit.

## 2.2. Neutral Atomic Hydrogen

Finally the fraction of total cold gas mass to total baryonic mass of a galaxy, the gas fraction, is $f_{gas} = (M_{molgas}+M_{HI}) / (M_*+M_{molgas}+M_{HI})$. Here $M_{HI}$ is the integrated atomic hydrogen mass of a galaxy. In massive *z*~0 SFGs the integrated atomic molecular hydrogen content dominates the total mass of the neutral ISM, $M_{HI}$~2-3 $M_{molgas}$ (e.g. Saintonge et al. 2011a and references therein, Catinella et al. 2010, 2013). The evolution of the atomic gas content of galaxies with redshift is relatively poorly known, since the HI 21cm emission line cannot be detected outside the local Universe with available technology. Bauermeister, Blitz and Ma (2010, and references therein) summarize the results coming from UV damped Lyman-α absorbers toward high-z QSOs, and conclude that HI columns likely do not vary strongly with redshift, while the molecular component strongly evolves with redshift (this paper, Daddi et al. 2010b, Tacconi et al. 2010, 2013, G15, Lagos et al. 2015a). For these reasons we make the approximation $\mu_{gas} \sim \mu_{molgas}$ and $f_{gas} \sim f_{molgas}$, which is valid at *z*>0.4 (see also section 4.2.1).

## 2.3 Determining Molecular Gas Masses: CO, Dust-SED and Dust-1mm Methods

There are currently three main avenues to obtain molecular gas masses:

1. The most common and well established method is the observation of a low-lying CO emission line (CO 1-0, 2-1, 3-2), and using its integrated line luminosity $L_{CO}$ and a conversion factor (or function) $\alpha_{CO}$ to convert $L_{CO}$ to molecular gas mass, in the regime where CO comes from optically thick, virialized clouds (e.g. Dickman, Snell & Schloerb 1986, Solomon et al. 1987, Bolatto et al. 2013);

2. More recently, high quality far-infrared/submillimeter dust emission SEDs have become available with the Herschel mission. From these SEDs, dust masses are inferred by fitting dust emissivity models (e.g. Draine & Li 2007). Then molecular gas masses are estimated assuming a gas-to-dust mass ratio (which can be a function of metallicity, see below, e.g. Leroy et al. 2011, Rémy–Ruyer et al. 2014, Eales et al. 2012, Magdis et al. 2011, 2012a, Magnelli et al. 2012a, Santini et al. 2014, Sargent et al. 2014, Béthermin et al. 2015, Schreiber et al. 2015, Berta et al. 2013, 2016, G15). Detailed explanations of this method are given in the above papers, and especially in G15 and Berta et al. (2016);
3. A single band measurement of the dust emission flux on the Rayleigh-Jeans side of the SED (at ~ 1mm) can be used to infer a dust/gas mass, if a single, constant dust temperature is a sufficiently good approximation. The long-wavelength dust emissivity can be estimated from a model, or calibrated from observations of sources in which gas masses are known from method 1 (e.g. Scoville et al. 2014, 2016, 2017).

All three methods have strengths and weaknesses (e.g. Bolatto et al. 2013, G15, Scoville et al. 2016, A.Weiss et al. in preparation). The 'CO method' is very well established and calibrated from observations in the local Universe (Bolatto et al. 2013), but at high-z the CO 1-0 line typically has to be replaced by a higher energy state line, which requires a calibration of the temperature and density dependent ratio $R_{1J}=T_1/T_J$, where $T_J$ is the beam averaged, Rayleigh-Jeans brightness temperature of the line J→J-1. Observations in low- and high-z SFGs suggest $R_{12}$~1.16-1.3, $R_{13}$~1.8 and $R_{14}$~2.4 (e.g. Weiss et al. 2007, Dannerbauer et al. 2009, Bothwell et al. 2013, Bolatto et al. 2015, Daddi et al. 2015).

With Herschel the FIR dust SED method (2) has been widely used for large samples between $z$=0 and 1, and for stacks between $z$=1 and 2.5. However, the FIR data covering the emission peak are luminosity weighted toward the warm, star forming dust component, and less sensitive to colder dust between star formation regions (Scoville et al. 2016, Carleton et al. 2017). This bias can lead to an underestimate of the ISM mass, especially at low-z, where only a fraction of the cold ISM is actively star forming. The effect will be less at high-z where the entire galaxy is globally unstable to star formation (Genzel et al. 2008, 2011, Elbaz et al. 2011). The Herschel-based references used in this study address this concern by fitting DL07 dust models that implement a distribution of dust temperatures. The rest wavelength coverage needed to constrain the colder dust has been studied by Draine et al. (2007), Magdis et al (2012a), and Berta et al. (2016). Another concern is that reaching down to the MS-line at $z$>1 cannot be done with single source detection photometry but requires stacking. We refer to the above papers for discussions of how stacking impacts the deduced masses. Analysis of the same or similar high redshift data sets can lead to significant differences (0.02 to 0.4 dex) in inferred dust masses, depending exactly on the assumptions and methodology (Santini et al. 2014, Béthermin et al. 2015, G15, Berta et al. 2016). The details of the different dust SED modeling methods are given in these papers.

Finally the '1mm'-method (3) is becoming increasingly popular as it is much more efficient than the CO method (factors ~5-10 in observing time at ALMA), if a constant dust temperature for the emitting dust grains at 1mm can be assumed ($T_{dust}$=25 K: see Scoville et al. 2016, 2017, A.Weiss et al. in preparation, for a full discussion). Alternatively, a second photometric measurement at shorter wavelength can constrain the dust temperature, but this can be costly in observing time (G15).

All three methods assume that zero-point calibrations established at $z$=0 are valid at higher redshifts without much change. Masses inferred from both CO and dust emission are sensitive to metallicity. In the case of CO, the conversion factor increases with decreasing metallicity $Z$, because CO is photo-dissociated to a larger depth in each cloud (Wolfire et al. 2010, Bolatto et al. 2013). Various metallicity dependent conversion functions have been proposed in the literature. As in G15,we adopt the geometric mean of the $\alpha_{CO}(Z)$ recipes of Bolatto et al. (2013) and Genzel et al. (2012) (equations 6 and 7 in G15)

$$\alpha_{\mathrm{CO\,J}} = 4.36\times R_{J1}\times\sqrt{0.67\times\exp(0.36\times10^{-(12+\log(O/H)-8.67)}\times10^{-1.27\times(12+\log(O/H)-8.67)}}$$

$$(M_{\odot}/(\mathrm{K\ km/s\ pc^2})) \qquad (2).$$

Here *logZ=12+log(O/H)* is the metallicity on the Pettini & Pagel (2004) scale. Molecular gas masses are then computed from equation (4) in G15. The recipes of Bolatto et al. and Genzel et al. are similar near solar metallicity but then deviate from each other below ~0.5 $Z_{\odot}$. In the sub-solar regime the exponential dependence of the Bolatto et al. recipe drives a much steeper increase of $\alpha$ than the power law in the Genzel et al. recipe. As a compromise we took an average of the recipes (the harmonic mean corresponds to the average in log-space). We discuss the impact of choosing individual prescriptions in the low-metallicity regime in section 4.2.3.

In the case of the dust methods (2) and (3) we assume that the dust-to-gas ratio is nearly linearly correlated with metallicity, at least for metallicities 12+log(O/H)>~8, as found in Leroy et al. (2011) and Rémy-Ruyer et al. (2014). We adopt the ratio of molecular gas to dust mass as

$$\delta_{gd} = \frac{M_{molgas}}{M_{dust}} = 10^{(+2-0.85\times(12+\log(O/H)-8.67))} \qquad (3).$$

Note that we deviate here from the assumption $\delta_{gd}$=const. in Scoville et al. (2016) but otherwise use their equation (16)[2].

For the few SFGs in this paper with estimates of gas phase metallicities from rest-frame optical strong line ratios, we determine individual estimates of *logZ=12 + log(O/H)*, adopting the Pettini and Pagel (2004) scale (e.g. Kewley & Ellison 2008 for a detailed discussion). However, for most of the SFGs in the CO and dust samples, such line ratios are not available and it is necessary to use the *mass-metallicity relation*, for the metallicity corrections discussed above. Following G15 we adopt

$$12+\log(O/H)_{PP04} = a - 0.087\times(\log M_* - b)^2, \text{ with}$$
$$a=8.74(0.06), \text{ and}$$
$$b=10.4(0.05) + 4.46\,(0.3)\times\log(1+z) - 1.78(0.4)\times(\log(1+z))^2 \qquad (4).$$

## 2.4 Summary of Previous Results

To set the scene, we summarize in Table 1 previous work on the molecular gas scaling relations, including several theoretical papers describing the results from hydro-simulations and semi-analytic work.

- *Redshift Dependence.* There is broad qualitative agreement in the literature that the depletion time scale is *about one Gyr*, and dropping by a factor of 2 to 4 between z=0 and 2.5 (Bigiel et al. 2008, Leroy et al. 2013, Saintonge et al. 2011b, 2013, 2016, Tacconi et al. 2010, 2013, Daddi et al. 2010b, Santini et al. 2014, Sargent et al. 2014, Genzel et al. 2010, G15). In comparing different estimates, those with a greater redshift-range are naturally preferable (see Appendix A, where we investigate the effects that limited parameter space and source statistics have on the inferred scaling relations). Since $\mu_{gas}=sSFR\times t_{depl}$ a depletion time that is slowly varying with redshifts means that molecular to stellar mass ratios nearly track *sSFR(z)*, and gas fractions increase strongly with redshift. The theoretical work also finds strong redshift evolution of increasing molecular (not atomic!) gas fractions with redshift (e.g. Lagos et al. 2012, 2015a, Genel et al. 2014, Popping et al. 2015).

[2]Note that the term $\Gamma_{RJ}/\Gamma_0$ in that equation has to be replaced by its inverse

- *Depletion time along and perpendicular to MS*. There is also broad qualitative agreement that $t_{depl}$ decreases as one steps upward in *sSFR* perpendicular to the MS line at a given *z* and $M_*$, as long as one considers a large enough range in $\delta MS$ (Saintonge et al. 2011b, 2012, Leroy et al. 2013, Tacconi et al., 2013, Huang & Kauffmann 2014, Sargent et al. 2014, G15). The theoretical work is in agreement with these findings (e.g. Lagos et al. 2012, 2015a, Genel et al. 2014, Popping et al. 2015). Depletion times are constant or increase slowly, stepping along the MS-line. Since $\mu_{gas}=sSFR\times t_{depl}$ this implies that gas fractions track the mass-dependence of *sSFR*, (Saintonge et al. 2011b).
- *Other parameters*. Studies of scalings on kpc scales within galaxies are so far only available at z~0, in particular through the HERACLES survey (Leroy et al. 2008, 2013, Bigiel et al. 2008, 2011). The HERACLES data do not exhibit strong intra-galactic parameter dependencies, with the exception of a significant drop of $t_{depl}$ at low galaxy masses, which likely reflects the impact of UV photo-dissociation in low-metallicity ISM clouds, and a resulting change in $\alpha_{CO}$ (Leroy et al. 2013). The HERACLES data do not show a significant dependence of $t_{depl}$ on (gas, stellar) density, or galactic radius, with the exception of a drop in the circum-nuclear regions. There is also no significant change in $t_{depl}$ (or its inverse, the star formation efficiency) between arm and inter-arm regions in M51, NGC628 and NGC6946 (Foyle et al. 2010). In contrast Huang & Kauffmann (2015) find a significant dependence on star formation and stellar surface density ($\log t_{depl}\sim -0.36\times\log\Sigma_{SFR} -0.5\times\log\Sigma_*$) from a different analysis of HERACLES, in combination with COLD GASS. From COLD GASS alone Saintonge et al. (2012, 2016) find that the depletion time may increase with stellar surface density, as the mass fraction of quenched bulges/spheroids increases. Expanding a subset of the HERACLES sample with HCN 1-0/CO 1-0 line ratios (as dense gas tracer), Usero et al. (2015) find that the L(IR)-to-L(HCN) ratio, thought to be closely related to the star formation efficiency of dense molecular gas, decreases systematically with these same parameters and is much lower near galaxy centers than in the outer regions of the galaxy disks. For fixed conversion factors, these results are incompatible with a simple model in which star formation depends only on the amount of gas mass above some density threshold (Lada et al. 2012).

Table 1 shows that despite the qualitative agreement mentioned above, uncertainties are large enough and methodologies sufficiently different to result in ambiguous and sometimes contradictory conclusions. The different methodologies for computing *SFRs* and stellar masses alone can lead to significant differences. At *z*=0, for instance, Bigiel et al. (2008) and Leroy et al. (2008, 2013) find $t_{depl}(MS)$~2-2.5 Gyrs vs. 1-1.5 Gyr from Saintonge et al. (2011b, 2012), emphasizing the importance of homogenized definitions and calibrations (benchmarking). At *z*=0, Bigiel et al. (2008, 2011) and Leroy et al. (2008, 2013) find a constant depletion time, with the exception of galactic nuclei, while Saintonge et al. (2011b, 2012) and Huang & Kauffmann (2014, 2015) emphasize that depletion time is correlated inversely with *sSFR*. The different conclusions could be caused in part by the narrower range of *sSFRs* covered in the HERACLES sample analyzed in Bigiel et al. (2008, 2011) and Leroy et al (2008, 2013), relative to COLDGASS. The derived values for the slope $d\log t_{depl}/d\log\delta MS$ vary from -0.2 to -0.7 (Saintonge et al. 2012. 2013, Santini et al. 2014, Sargent et al. 2014, G15, Scoville et al. 2016, 2017). Simulations and semi-analytic work continue to find somewhat lower (by a factor of 1.5-2) *SFRs* and gas fractions at high-z than implied by most observations (e.g. Davé et al. 2011, 2012, Lagos et al. 2015a, Genel et al. 2014).

Another important example is the interpretation of the MS itself. A massive ($M_*$~$10^{11}$ $M_\odot$) galaxy on the mid-plane of the MS at z~2.3 has an *SFR* of ~200 $M_\odot yr^{-1}$. The same *SFR* is reached at *z*=0 only for (ultra)-luminous infrared galaxies ((U)LIRGs), which are placed well above the MS by major merger triggered starbursts (Sanders & Mirabel 1996). This does not mean that all z=2.3 MS SFGs are mergers, however. On the contrary, the advent of Herschel SEDs (see Elbaz et al. 2011, Nordon et al. 2012), and the firm establishment of the MS-picture discussed in Section 1 favor an explanation where the increase in *SFRs* on the MS in equation 1 implies that normal star forming disk galaxies at high-z are more gas-rich. While the former ('starburst') interpretation had dominated earlier work, the more recent results now favor the 'equilibrium growth model' (section 1.1, see Elbaz et al. 2011, Nordon et al. 2012). Santini et al. (2014) and Scoville et al. (2016) find only a moderate slope of $\mu_{gas}$ with z, suggesting the need for a more efficient star formation process at high-z, perhaps driven by the increased merger rates. In contrast G15 find that $\mu_{gas}$ changes rapidly with redshift, tracking *sSFR(z)* and favoring a single dominant star formation process on the MS at all redshifts between *z*=0 and 2.5. Daddi et al. (2010a,b), Genzel et al. (2010), Magdis et al. (2012a) and Sargent et al. (2014) all applied a Galactic CO conversion factor near the MS, but a much smaller one above the MS, motivated by observations of local ULIRGs (Downes & Solomon 1998). This resulted in the proposal that galactic star formation is 'bi-modal', with large depletion time scales (low efficiency) at $\delta MS$~1 and short depletion time scales (high efficiency) at $\delta MS$ >>1, with a fairly sudden transition in between. G15 did not see much difference in the dependence of $t_{depl}$ on $\delta MS$ between CO and dust-based data, suggesting that strong $\alpha_{CO}(\delta MS)$ changes are probably not justified, at least at high-z. The work of Scoville et al. (2016, 2017) independently reaches the same conclusion.

As we show below, many of these differences are caused by output parameter estimation of often modest data sets across limited input variable ranges, in addition to the systematic differences in calibrations of the input variables entering the analyses. By substantially increasing these variable ranges, the total number of data points and including different calibration methods, we expect significantly more robust results.

## 3. Data Sets Entering this Analysis

The data used in this work comprise a total of 1444 measurements of molecular mass, spanning a wide redshift range *z*=0 to 4.4, over three orders of magnitude in stellar mass from log($M_*$/$M_\odot$)=9-11.9, and with *SFRs* from $10^{-2}$ to $10^2$ times those on the MS-reference line. This data set is about 40% larger than that of G15, because of more CO measurements (mainly from PHIBSS2, Freundlich et al. 2018, PHIBSS2 team in preparation), more FIR SED dust measurements (Santini et al. 2014, Bethermin et al. 2015) and, most importantly, because of a set of 1mm dust measurements (mostly from Scoville et al. 2016). This allows an update of G15 with somewhat better statistics, but more importantly, a thorough comparison with the same analysis of the different methods of determining molecular hydrogen masses and columns.

We omit 136 dust measurements between *z*=0.1 and 0.4 for analysis of the redshift dependence of $\mu_{gas}$, since these may be affected by dust in atomic gas (section 2.2 above and also sections 3.2 and 3.3 below), such that the dust and inferred gas masses in this redshift range overestimate the true molecular masses (e.g. Draine et al. 2007, Kennicutt et al. 2011, Dale et al. 2012, Eales et al. 2012, Sandstrom et al. 2013). These lower z dust measurements were included in the results presented in G15, where the effect was a shallower falloff of $\mu_{gas}$ with z relative to the relation using CO data points alone (Table 4 of G15). As we show below and in Table 3a, omitting or including the low-z dust points has little effect on the slope of the $t_{depl}$ relation with redshift.

The final data set used in our global analysis of $\mu_{gas}$ thus has 1309 measurements. Of those, 667 come from CO line flux measurements, 512 come from dust masses inferred from stacking Herschel FIR spectra, and 130, of which 22 are stacks, from dust masses inferred from broadband 1mm photometry. Table 2 shows the key quantities used in the analysis for 7 CO and 7 dust data points. The full table will be available online at http://www.iram.fr/~phibss2/Home.html. In order not to bias the trends, we include >3σ individual detections in single dish and interferometric spectroscopy (where we have a spectroscopic redshift as additional information), >4σ for individual continuum detections and >4σ for stacks, noting that values for individual galaxies at the lower confidence levels are very uncertain. For the individual detections of SFGs ($\delta MS$>0.1) the fraction of targets detected at ≥3σ, ≥4σ and ≥5σ varies between various samples and measurement methods. For the xCOLD GASS surveys (Saintonge et al. 2011a,b, 2016, 2017) these fractions are 0.99, 0.93 and 0.86 for star forming galaxies. For the PHIBSS 1 and 2 surveys (Tacconi et al. 2013, Freundlich et al. 2018, and PHIBSS2 team in preparation) the fractions are 0.92, 0.79 and 0.64, respectively. For the FIR-SED stacks the detection fraction is high, and mass errors are driven by SED modelling assumptions and methods (Berta et al. 2016). For the 1mm photometry samples (Scoville et al. 2016, DeCarli et al. 2016, Dunlop et al. 2017, Barro et al. 2016, Tadaki et al. 2017, S. Lilly, priv.comm.) the detection fractions are about 0.65, 0.53 and 0.4 for ≥3, ≥4 and ≥5σ significance. For the stacks of Scoville et al. (2016) the detection fractions are 0.86 for ≥5σ. Detection rates below $\delta MS$<0.1 plummet, and are not included in this analysis. Because of the generally high detection rates of the different methodologies we did not attempt to apply completeness or significance corrections.

### 3.1 CO observations

We collected 667 CO detections of SFGs from a number of molecular surveys with CO 1-0, 2-1, 3-2 (and in two cases 4-3) rotational line emission. These data cover the redshift range from $z$=0 to 4.0, the stellar mass range of $M_*=10^{9.0}$ to $10^{11.8}$ $M_\odot$ ($M_* < 10^{10}$ $M_\odot$ for $z$=0 only), and at a given redshift and stellar mass, *SFRs* from about $10^{-1}$ to $10^2$ times the MS-SFR. We include

1. 216 detections of CO 1-0 emission above and below the main sequence between $z$=0.025-0.05 from the final xCOLD GASS survey with the IRAM 30m telescope (Saintonge et al. 2011a,b, 2016, 2017), and the single COLD GASS stack detection of galaxies much below the main-sequence. We also include 89 detections of the low mass extension of XCOLD GASS ($log(M_*/M_\odot)$=9.0-10.0, Saintonge et al. 2017). We note that the star formation rates in COLD GASS have been updated from earlier UV-/optical SED fitting (Saintonge et al. 2011a) with mid-IR star formation rates from WISE (Saintonge et al. 2016, Huang & Kauffmann 2014).
2. 90 CO 1-0 detections with the IRAM 30m of $z$=0.002-0.09 luminous and ultra-luminous IR-galaxies (LIRGs and ULIRGs) from the GOALS survey (Armus et al. 2009), from the work of Gao & Solomon (2004), Gracia-Carpio et al. (2008, 2009, and priv. comm.), and Garcia-Burillo et al. (2012).
3. 31 CO 1-0 or 3-2 detections of above main-sequence SFGs between $z$=0.06 and 0.5 with the CARMA millimeter array from the EGNOG survey (Bauermeister et al. 2013).
4. 14 CO 2-1 or 3-2 detections at $z$=0.6-0.9 and 18 CO 1-0 detections at $z$=0.2-0.58 (significantly above-main-sequence) ULIRGs with the IRAM 30m telescope from Combes et al. (2011, 2013).
5. 51 detections of CO 3-2 emission in main-sequence SFGs in two redshift slices at $z$=1-1.5 and $z$=2-2.5 as part of the PHIBSS1 survey with the IRAM PdBI (now NOEMA; Tacconi et al. 2010, 2013).
6. 97 detections of CO 2-1 or 3-2 in main sequence SFGs between $z$=0.5 and 2.7, as part of the PHIBSS2 survey with the updated IRAM NOEMA interferometer (G15, Freundlich et al. 2018, and PHIBSS2 team, in preparation).

7. 9 CO 2-1 or 3-2 detections of near main-sequence SFGs between *z*=0.5 and 3.2 from Daddi et al. (2010a) and Magdis et al. (2012b), obtained with the IRAM PdBI.

8. 6 CO 2-1 detections of *z*=1-1.2 main-sequence SFGs selected from the Herschel-PEP survey (Lutz et al. 2011), obtained with the IRAM PdBI (Magnelli et al. 2012b).

9. 19 CO 2-1, 3-2 or 4-3 detections of above main-sequence submillimeter galaxies (SMGs) between *z*=1.2 and 3.4, obtained with the IRAM PdBI by Greve et al. (2005), Tacconi et al. (2006, 2008) and Bothwell et al. (2013).

10. 8 CO 3-2 detections of *z*=1.4 to 3.2 lensed main-sequence SFGs obtained with the IRAM PdBI (Saintonge et al. 2013, and references therein).

11. 10 ALMA CO 3-2 and 2-1 detections between *z*=1 and 2.5 (Genzel et al. in preparation, De Carli et al. 2016), and

12. 7 CO 2-1 and 3-2 detections of *z*=1.4 to 2.2 'outliers' above the MS, with ALMA and NOEMA (Silverman et al. 2015).

### 3.2 Dust observations

We take from the literature thermal continuum dust observations from Herschel and ALMA, which have molecular gas masses estimated from methods (2) and (3) described above. We include

1. 512 (Magnelli et al. 2014, Berta et al. 2016, G15) and 121 (Santini et al. 2014) stacks of *z*=0.1-1.9 SFGs with deep Herschel PACS/SPIRE spectro-photometry in the COSMOS and GOODS (N/S) fields as part of the PEP (Lutz et al. 2011) and HerMES (Oliver et al. 2012) surveys, and 15 stacks of *z*=0.4 -3.8 SFGs in COSMOS, again with deep PACS/SPIRE photometry, plus additional short- and long-wavelength coverage with Spitzer, LABOCA and AzTEC (Béthermin et al. 2015). In all cases we adopted dust masses from these references, and converted to gas mass as described in Sect 2.1. For the final analysis, we removed 136 stacks at *z*<0.4 that are likely significantly affected by dust in HI gas, **as discussed above**.

2. 102 detections and 21 stacks of *z*=0.9-4.4 of 850-1300 μm continuum fluxes from recent ALMA observations (Scoville et al. 2016, Dunlop et al. 2017, DeCarli et al. 2016, Barro et al. 2016, Tadaki et al. 2015, 2017, S. Lilly et al. 2018, in preparation), using the methodology and calibration proposed by Scoville et al. (2016) with $T_{dust}$=25K=const., but correcting for the metallicity dependence of gas to dust ratios as discussed above. In the case of Scoville et al. (2016) we included the 72 individual detections with significance ≥4σ and redshifts verified by additional data (S.Wuyts, priv.comm.). For the 21 stacks in the *z*~1 and ~2 bands presented by Scoville et al. (2016), 18 (0.86) have a significance >4σ and were included. We also included the average of 1mm photometry detections of 45 SFGs between *z*=2.8 and 3.8 as published in Schinnerer et al. (2016). Finally we included 6 1mm dust continuum detections between *z*=1.2 and 2.3 with the IRAM NOEMA interferometer (from the PHIBSS2 survey), for a total of 108 individual detections and 22 stacks in the 1mm photometry technique.

### 3.3 Benchmarking

Our basic approach is that the core, near main sequence, CO or dust data sets are "benchmark" sub-samples of large panchromatic (UV/optical/infrared/radio) imaging surveys, preferably with spectroscopic redshifts, and with well-established and relatively homogeneous stellar and star formation properties. We note that we eliminated some of the *z*~4 data points of Scoville et al. (2016), since the grism redshifts from 3D HST and the literature were discrepant with the photometric redshifts used in that paper. The xCOLD GASS sample is mass-selected from the SDSS (Saintonge et al. 2011a, b, Saintonge et al. 2016, 2017). PHIBSS 1 & 2 and the Herschel and ALMA dust samples are selected from deep rest-frame UV-/optical imaging surveys in the EGS (Davis et al. 2007, Newman et al. 2013, Cooper et al. 2012), GOODS N/S (Giavalisco et al. 2004, Berta et al. 2010) and COSMOS fields (Scoville et al. 2007, Lilly et al. 2007, 2009), including the recent CANDELS J- and H-band HST imaging (Grogin et al. 2011, Koekemoer et al. 2011) and 3D-HST grism spectroscopy (Brammer et al. 2012, Skelton et al. 2014, Momcheva et al. 2016). We also include galaxies from the Deep-3a survey (Kong et al. 2006) and the BX/BM surveys of Steidel et al. (2004) and Adelberger et al. (2004). We have supplemented these core samples with smaller datasets addressing outliers above the MS, mainly starburst sources (LIRGS, ULIRGs, submm-galaxies etc.) as described in sections 3.1 and 3.2.

### 3.4 Stellar Masses and Star Formation Rates

With these selections it is possible to place the basic galaxy parameters: stellar masses, *SFRs*, effective radii in the rest-frame optical, on a common '*ladder*' system (see Wuyts et al. 2011a,b and Saintonge et al. 2011a, 2016 for details), where *SFRs* are based on FIR emission, MIR emission, and UV to NIR SED fitting in decreasing preference. The impact of *the availability or absence of mid-/far-infrared photometry is quite important*, as the comparison of the left and right panels of Figure 1 shows. *SFRs* based only on optical-/UV-SED analysis tend to underestimate the total *SFRs*, which is especially relevant at high redshifts, low stellar masses and below the MS. Wherever necessary and possible, we adjusted the stellar masses and *SFRs* from the literature to the same assumptions. Typical fractional stellar mass uncertainties (including systematic errors) are ±0.13 dex on the MS and ±0.2 dex for outliers, star formation rate uncertainties are ±0.2 dex for Herschel-Spitzer detected galaxies, and ±0.25 dex for SED-inferred SFRs, or starbursts, and gas mass uncertainties are ±0.23 dex (G15). Note that throughout the paper we define stellar mass as the "observed" mass ("live" stars plus remnants), after mass loss from stars. This is about 0.15…0.2 dex smaller than the integral of the SFR over time. The redshift-*sSFR*-$M_*$ coverage of the various samples is shown in Figure 2, with the different symbols denoting the various surveys mentioned in our listing above.

### 3.5 How well do our CO and Dust Samples Represent the Parent Samples?

Figures 2 & 3 give the distribution of our data in the *logsSFR-log(1+z)* plane separately for the three methods, in the *logsSFR-log*$M_*$ (at z=0 for the CO method), and in the $\log R_e(5000\,Å)$-$\log(1+z)$ plane. We remind the reader that it is not realistic to construct an unbiased gas sample, whose distribution function in these planes is proportional to the distribution function of the parent sample. Rather the question is how broad and unbiased the coverage in each of the relevant parameters is, from which we then can attempt to determine scaling relations.

Overall the parameter coverage of the combined sample is quite broad: redshift *z*=0 to 4.4, stellar mass from $\log(M_*/M_\odot)$=9-11.9, *SFRs* from $10^{-2}$ to $10^{2}$ times those on the MS-reference line, and sizes 0.25 to 2 times the typical size at a given mass and redshift. The best coverage occurs at *z*=0 with the xCOLD GASS, and LIRG/ULIRG CO surveys. Unfortunately we have no access to equivalent surveys in the dust tracers of the molecular ISM, due to the strong contribution from dust in the atomic medium (e.g. Draine et al. 2007, Kennicutt et al. 2011, Dale et al. 2012, Eales et al. 2012, Sandstrom et al. 2013). The low mass galaxy coverage from $\log(M_*/M_\odot)$=9-10 is also exclusively from the xCOLD GASS CO survey. In the mid-z range (*z*=0.5-2.5) we have the best comparison of the three independent methods, and with several independent analysis methods of the SED technique. Owing to sensitivity limits, the overall distribution in *z-sSFR* space at all z is somewhat biased to SFGs on and above the main sequence and at higher stellar masses ($<\log\delta MS>$=0.2-0.34). However, the recent extensive surveys at the IRAM telescopes at *z*~0.03 (xCOLD GASS), *z*~0.7 (PHIBSS2), *z*~1.2 (PHIBSS1+2) and *z*=2.2 (PHIBSS1+2) now establish good coverage of massive SFGs above and below the main-sequence line (Figure 2). In comparison, the other large survey of the molecular gas properties in high-z SFGs from 1mm ALMA dust photometry (Scoville et al. 2016, 2017) has $<\log\delta MS>$=0.5-1 between z=1-3 and thus is even more strongly biased to above MS galaxies.

As mentioned above, with the exception of xCOLD GASS (Saintonge et al. 2017), none of the other samples reaches below $M_*\leq 10^{10}M_\odot$. This is in part because of sensitivity limitations for the higher redshift surveys, but more importantly it is also *per design*, as in this mass range the (line or continuum) luminosity per ISM mass decreases rapidly due to the metallicity dependence of $\alpha_{CO}$ and $\delta_{gd}$ (section 2.1). Given these limitations, we discuss in sections 4.2.3 and 4.3 what we can infer from xCOLD GASS for the stellar mass dependence of the scaling relations for $M_*\leq 10^{10}M_\odot$, but we caution the reader that the relations at the low mass end are not represented by galaxies with z>0.05.

# 4. Results

## 4.1 Separation of Variables: $t_{depl}(z, \delta MS, \delta M_*, \delta R_e)$

In the framework of the MS-prescription (equation (1)) and following the analysis in G15 (section 2) and other papers, our Ansatz is to separate the parameter dependencies of $t_{depl}$ as products of power laws, first in redshift *(1+z)*, next at a given redshift above and below the MS at a fixed stellar mass, ($\delta MS=sSFR/sSFR(MS,z,M_*)$), and then along the MS ($\delta M_*= M_*/5\times10^{10}M_\odot$, corrected to fiducial stellar mass of $5\times10^{10}$ $M_\odot$). Finally we investigate the residuals as a function of effective radius (half-light radius in rest-frame optical (5000 Å) $R_e$), relative to the average radius of the star forming population, $R_{e0}=8.9\ kpc\ (1+z)^{-0.75}(M_*/5x10^{10}M_\odot)^{0.23}$ (van der Wel 2014), such that $\delta R=R_e/R_{e0}$. This means that

$$\log(t_{depl}(z,sSFR,M_*,R_e))= \\ A_t + B_t\times\log(1+z) + C_t\times\log\left(\delta MS\right) \\ + D_t\times\log\delta M_* + E_t\times\log\left(\delta R\right) \qquad (5).$$

Separation of variables requires that the parameters $C_t$, $D_t$ and $E_t$ should not depend significantly on redshift. We show in section 4.2.3 and 4.2.4 that $D_t$ and $E_t$ are indeed close to zero and can be neglected to first order.

To explore the redshift- dependence of $C_t$, we first split the independent data sets of each of the three methods (CO, dust-FIR, dust-1mm photometry) into six redshift bins (*z*=0-0.1, 0.1-0.5, 0.5-0.9, 0.9-1.6, 1.6-2.5 and 2.5-4.4). In each of the redshift bins and separately for each of the three methods, we fit for $A_t'=A_t+B_t\times log(1+z)$ and $C_t$. In the literature there are three independent analyses of the FIR/submm dust-SEDs from Herschel (+Spitzer, ground-based: Berta et al. 2016 and G15 (using fluxes from Magnelli et al. 2014), Santini et al. 2014 and Béthermin et al. 2015), in which $t_{depl}$ and $\mu_{gas}$ were determined in stacks, as a function of *z*, $\delta MS$ and $\delta M_*$. We analyzed each of these data sets separately. The right panel of Figure 4 summarizes the inferred slopes $C_t$. There is no overall significant redshift trend of $C_t$. The distribution of individual values of $C_t$ around the best fit, error-weighted average (-0.44) has a scatter of ±0.22 dex, somewhat larger than the median uncertainty of the individual data points (±0.15 dex). If data points of Béthermin et al. (2015) are not considered, that scatter further decreases to ±0.16 dex, suggesting that the data can be described mostly with scatter around a *constant, redshift independent slope, in excellent* *agreement with the Separation Ansatz*. The $C_t$-values inferred from the Magnelli et al. (2014) and Berta et al. (2016) points (with very similar input data) differ on average by $\Delta C_t$~0.36 dex. This suggests that the scatter is significantly affected by systematics in the analyses. We adopt the overall best slope of $C_t$= -0.44, which includes all the datasets listed in Section 3 (see Table 3a).

Once the scaling relations for $t_{depl}$ are determined, it is in principle straightforward to determine the equivalent relations for molecular gas to stellar mass ration, $\mu_{gas}$, from the combination of equations (1) and (5),

$$\begin{aligned}\log(\mu\ (z, sSFR, M_*, R_e)) &= \log(M_{molgas} / M_*) \\ &= \log t_{depl} + \log(sSFR) = \\ &= \log t_{depl} + \log(sSFR(MS, z, M_*)) + \log\left(\frac{sSFR}{sSFR(MS, z, M_*)}\right) \\ &= A_\mu + B_\mu \times \left(\log(1+z) - F_\mu\right)^\beta + C_\mu \times \log\left(\delta MS\right) \\ &\quad + D_\mu \times \log(\delta M_*) + E_\mu \times \log\left(\delta R\right) \qquad (6).\end{aligned}$$

Unfortunately, the slope of the MS-line (*dsSFR(MS)/dlog(1+z)*) in S14 and also in W14 varies quite strongly with redshift (from +3.6 at $z$~0-1 to +1.2 at $z$>2.2, at $\log M_*$~10.8) and mass. Linear functions in log(1+$z$) and log $\delta M_*$, based on the scaling relations obtained in equations 1 and 5, thus are not sufficient. To capture the slope variations we introduced two more parameters, $F_\mu$ and β (=2), as shown in equation 6. Because of the slope-variations of *sSFR*(MS) in *z*, fitting of the data is quite sensitive to the range and distribution of data points (especially for the pure power-law case β=0 and $F_\mu$=0). For this reason we also fit the data by first binning in z and then giving all z-bins equal weight for the determination of the parameters $A_\mu$, $B_\mu$, $C_\mu$, and $F_\mu$.

### 4.2 New Results for Depletion Time Scaling Relations

In the following we use two approaches in parallel. First, to visualize specific trends in Figures 4 and 5, we use data averages/medians of typically 20-100 individual measurements, separately for each technique. Binning in this way elucidates possible deviations from the assumed basic power law parameterizations introduced in equations 5 and 6. Second, for quantitative fitting of the data, we employ multi-parameter linear regression fitting in 3-, 4- or 5-space of logarithmic variables (log(*1+z)*,$\log^2$*(1+z)*, log$\delta MS$, log$\delta M_*$, log$\delta R_e$) using all data points individually, weighted by the inverse square of their uncertainties. For the specific fitting of *log(1+z)* vs *log*$t_{depl}$ *or log*$\mu_{gas}$, we explored how the fits changed when we gave equal weight to each of the 6 redshift bins used in the fit. This is important for establishing the best overall *log(1+z)*-log $\boldsymbol{\mu_{gas}}$ scaling relation (equation 6) in the presence of the non-linear fitting function (β=2). The equal bias for different *z*-bins removes the otherwise overly strong weight of the large number of $z$~0 CO data points. The results of the fitting exercises are reported in Table 3a, including a recommended, overall '*best*' set of fit parameters (bold face). We determined the uncertainties in Table 3a by splitting the sample randomly into two halves, fitting the parameters, and then repeating the splitting and fitting to establish the range of uncertainties by bootstrapping. As a second check, we also did the fitting procedures after eliminating one or more of the smaller datasets (jackknifing), thus checking for systematics of the individual data sets (See also Appendix A).

### 4.2.1 Redshift dependence of depletion time scale

The most striking impression of the left panel of Figure 4 is that the redshift trends of the different methods sets are similar if one refers *to the depletion time at the MS-reference line*[3] ($B_t(\delta MS=1)=d\log t_{depl}(MS)/d\log(1+z)\sim-0.4\ldots-1.0$). The one exception are dust measurements in the lowest redshift range ($z$=0.1-0.4). The Santini et al. (2014) data indicate a fairly sharp upward trend of the depletion time scales with redshift at $z$<0.4. We suspect that this change in slope comes from dust in the atomic gas component of the galaxies becoming an important contribution to the total observed dust mass, since $M_{HI}$ and $M_{molgas}$ are likely comparable there (e.g. Catinella et al. 2010, 2013, Maddox et al. 2015, Saintonge et al. 2016). For this reason we eliminated these low-$z$ dust data in our further analysis, as described earlier in the paper.

There are obvious zero-point differences between different methods and data sets. We solved for these zero points by fitting a slope $B_t$=-0.6 line to each, determined the zero-point, and then computed each offset from the best-fitting common zero point ($A_t\sim0.09$). These zero-point corrections (Table 3a) are then applied to all data for further analysis. The bottom left panel of Figure 5 shows that the scatter around the best fitting line decreases from ±0.15 dex before, to ±0.066 dex after this correction. The resulting scaling of $t_{depl}(MS)$ with redshift becomes reasonably tight, with a *fairly shallow redshift dependence*, $t_{depl}(MS)\sim(1+z)^{-0.62\pm0.13}$.

The dust data appear to have a steeper slope $B_t$ than the CO data. This was already noted by G15 for the Magnelli/Berta data, and is confirmed by the Bethermin et al. (2015) and Santini et al. (2017) measurements, and by our 1mm data. However, uncertainties of $B_t$ in these dust data are large, likely because of their smaller z-coverage, and the difference to the CO data is probably not significant (Table 3a). From a 1mm dust sample of 708 SFGs between z=0.3 and 4 Scoville et al. (2017) recently reported $B_t$=-1.05 (±0.05), again steeper than our CO data, with a significance of ~3σ. The value and uncertainty of the z=0 CO data point is probably critical for understanding these differences. If the COLD GASS value of $\langle t_{depl}\rangle_{MS}$=1.1 Gyr is replaced by the HERACLES value (2-2.5 Gyrs), the CO slope changes to -0.8 (see Tacconi et al. 2013). Excluding the z=0 CO point still results in a shallow slope (-0.43) but with a 1σ uncertainty of ±0.28.

### 4.2.2 Variations of depletion time scale above and below the MS

Next we remove the average redshift dependence and consider the residual variations of depletion time as a function of $\delta MS$ in the left top panel of Figure 5. The data are well described with a single power law of slope -0.44 (±0.03), over a remarkably large range from $\log\delta MS$=-1 to +2 around the MS, from the 'green valley' to the regime of extreme outlier starbursts. The data are accurate enough to look for empirical deviations from a single power law description by inspecting the binned averages in the top left panel of Figure 5. While there might be a tendency for a flattening of the relation between $\log\delta MS$=0 to +1 in the CO data (more constant depletion time), and a steepening further out, these deviations are everywhere less than ±0.1dex. We conclude that a *single power law* ($t_{depl}\sim(sSFR/sSFR(\mathrm{MS}))^{-0.44\pm0.03}$ describes all data from the different methods. Magdis et al. (2012a) and Sargent et al. (2014) have proposed that there is a fairly sudden decrease of $\alpha_{CO}$ by -0.5 to -0.7 dex between $\delta MS$=0 and +0.6, motivated by the findings of Downes & Solomon (1998) in local ULIRGs. Our analysis does not support the presence of such a large change in $\alpha_{CO}$, confirming the findings of G15, and those of Scoville et al. (2016, 2017), now with the dust-1mm technique added as another independent anchor in the argument.

---

[3] because of the significant *sSFR*-dependence of the depletion time scale it is very important to not just compute an average or median depletion time measurements in a given z-bin, especially if the selection function is biased towards $\delta MS$>1, as tends to be the case for higher z. Instead we determine the $t_{depl}(\delta MS=1)$ value in each z-bin by fitting to all data in that bin a straight line with slope -0.44 in the $\log t_{depl}$-$\log\delta MS$ plane and solve for the zero-point value $A_t'(z)$.

### 4.2.3 Mass dependence

In the third step we analyze the residuals of depletion time along the MS, as a function of stellar mass (Figure 5 bottom right panel), once both *z* and *SFR*-dependences are removed. Over the mass range covered by all three techniques (log($M_*$/M$_\odot$=10-11.6) the depletion time scale data do not depend significantly on mass, and the relation is flat within ±0.08 dex (2σ) in the slope. If the lower mass SFGs in the CO xCOLD GASS survey are added (Saintonge et al. 2017, $log(M_*/M_\odot)$=9-10.2), or more weight is given to the xCOLD GASS survey as a whole, the slope increases slightly ($D_t$ ~0.05 to 0.1), but the trend remains marginal.

The data are better described by a second-order relation with curvature: $\log t_{depl}=0.05_{0.03} - 0.17_{0.06}\times(log(M_*)-10.8_{0.2})^2$. However, both the steepening of the relation and the potential curvature vary strongly on the metallicity-dependence of $\alpha_{CO}$. To demonstrate this, we show four different recipes for that dependence in the bottom right panel of Figure 5, from no correction (strongest negative deviations), to the Accurso et al. (2017) conversion function, which almost flattens the relationship. It thus remains uncertain whether these second order depletion time variations with mass are intrinsic, or whether they are indicative of $\alpha_{CO}$-metallicity dependences that are not captured in equation (2). On balance we recommend a $\log t_{depl}$-$logM_*$ dependence with a *flat or very shallow slope* ($D_t$~0-0.09).

### 4.2.4 Size dependence

Finally we looked for any size or surface density dependences, which have not yet been explored in previous work due to the limited sample sizes. Such dependences could well be related to the dependence of $t_{depl}$ *on sSFR*. Wuyts et al. (2016) found that the baryon fraction in SFGs is strongly correlated with surface density (of baryons, or stars). The residual size dependence of $t_{depl}$ is shown in the upper right panel of Figure 5. We plot the residuals, after correcting for *z, sSFR* and $M_*$-dependencies, as a function of rest-frame optical size. We derive the sizes from exponential fits to the observed H-band or R-band emission, and then correct to rest-frame 5000 Å effective radii according to the prescription in van der Wel (2014; their equation (2)), after removing the mean population trends as a function of redshift and stellar mass (van der Wel et al. 2014). In the local Universe optical continuum and CO sizes of main sequence galaxies correlate well empirically (e.g. Young & Scoville 1991, Leroy et al 2009, 2013), and the first spatially resolved CO sizes in large *z*~1-2 disks seem to support this assumption as well (Tacconi et al. 2013; M. Lippa et al. 2018 in prep.). The distribution in the upper right panel of Figure 5 is flat or marginally increasing ($E_t$=0.11±0.1).

It is premature to conclude that there is no surface density dependence of $t_{depl}$. High-z disks in the mass range probed by the majority of the samples studied here are highly dusty (Wuyts et al. 2011b). *Optical sizes may not represent the true molecular/dust effective radii*. Recent 1mm dust/CO imaging for a fraction of massive (>$10^{11}$ M$_\odot$) *z*~1-2.5 SFGs show the presence of compact, centrally concentrated dust/gas concentrations, with radii less than those measured in the rest-frame optical (Tadaki et al. 2015, 2017; Barro et al. 2016, Tacconi et al. 2013, M. Lippa et al. in preparation).

Another concern is that galaxy parameters such as specific star formation rate, mass and size may be inter-correlated and that our first removing the *z, δMS* and $M_*$ dependencies before studying the dependence on $R_e$ or surface density, may be misleading. If the intrinsic correlation of $t_{depl}$ is with $\Sigma_{SFR}$, one should study the correlation with that parameter directly. For the 1309 galaxies included here, this yields a relation $log\ t_{depl}$ (Gyr$^{-1}$) $=0.51_{0.06} - 0.26_{0.02}\times$ $log\ \Sigma_{SFR}$ (M$_\odot$yr$^{-1}$ kpc$^{-2}$). So there is significant correlation (in contrast to the residual-removed $R_e$-relation). However, the residual $\chi_r^2$ is 2.9, much worse than the other $\chi_r^2$ values in Table 1 (0.5-1.2). From a statistical point of view the hypothesis that the *z* and *δMS*-dependencies of $t_{depl}$ (and $\mu_{gas}$) merely encapsulate surface-density dependencies, thus is not supported. Finally, any residual systematic or random errors left from removing the z, δMS and mass could increase the scatter and mask a weak dependence on size.

In summary, considering the binned data, each of which is an average of 20 to 60 individual detections or stacks, all individual averages corrected for zero-point offsets, scatter with ±0.085 dex around a linear fit line. Individual depletion time measurements have a median 1σ uncertainty of ±0.26 dex (combined statistical (SNR) and systematic (calibration etc.) uncertainties). For purely white noise this should integrate down to ±0.047 dex per average of ~30 points. This means that the simple fitting approach proposed in equation (5) describes the 1309 data points used in the fit at better than ~2σ. We conclude that any additional systematics due to the input assumptions of the different molecular mass determination methods, or any undiscovered hidden correlations, are modest and the scaling relations describe the data at the ±0.1 dex level, or better.

### 4.3 New Results for Gas Fraction Scaling Relations

We now repeat the same exercise for the molecular gas to stellar mass ratio, $\mu_{gas} = M_{molgas}/M_*$, using equation 6 as the basis of our fitting. Figure 6 and Table 3b show the results. We first fit for the dependence of $\mu_{gas}$ as a function of redshift normalized to the MS-reference line of S14, applying the zero-point corrections to the different data sets. The result is shown in the bottom left panel of Figure 6. The decline in $\mu_{gas}$ over cosmic time is consistent with the results found previously in the literature based on fewer data points (see Table 1). We find that fitting the redshift-dependence of $\mu_{gas}$ across the entire range requires a curved, second order function in log $(1+z)$-log $\mu_{gas}$-space, and the best-fit with β=2 is given in Table 3b and clearly improves the quality of the fit over the simple power-law. This second order fitting function should not be surprising as it merely reflects the shape of *sSFR* in equation (1), and demonstrates that *sSFR and* $\mu_{gas}$ *of MS galaxies track each other*, as found previously by Tacconi et al. (2010, 2013) and others. Because of this curvature of $\mu_{gas}$ with local slope getting flatter the higher the mean *z*, fitting subsets of data with only partial redshift coverage will reflect this systematic flattening, for instance when only fitting the FIR dust or 1mm dust data that do not have data at $z<0.4$. We also show the mean total gas fraction ($HI+H_2$) at $z=0$ for a mean stellar mass of $\log(M_*/M_\odot)=10.7$, taken from Saintonge et al. (2011a). As mentioned above, the dust data of Santini et al. (2014), G15 and Berta et al. (2016) in the lowest redshift bin ($z\sim0.1$) are located above best fit relation, suggesting that dust from the atomic medium in the $z=0$ data of Saintonge et al. 2011a and 2017, could be becoming significant (section 3 and 3.2). In this interpretation $z\sim0.1$-0.3 would mark the transition between a primarily atomic-ISM and a primarily molecular-ISM. The mass contribution of the atomic component is approximately constant with redshift (Bauermeister et al. 2010), while the molecular component is strongly evolving with redshift (see also Lagos et al. 2011, 2015a).

With the redshift evolution established, we now fit for the dependence with *sSFR* (*δMS*) perpendicular to the MS. The top left panel of Figure 6 confirms the tight correlation found previously (e.g. Saintonge et al. 2012, 2016, Tacconi et al. 2013, Sargent et al. 2014, G15) that *gas fractions increase with δMS* ($\mu_{gas}\sim\delta MS^{0.52}$) . The increase in *sSFR* above the main sequence is then a *combination of an increase in the available gas for star formation and an increase in star formation efficiency* (decrease in depletion time as described in the previous section).

Next, we fit for the mass dependence after subtracting the dependences on redshift and *sSFR*, and show the result in the bottom right panel of Figure 6. The plot clearly shows the drop in $\mu_{gas}$ at high $M_*$ as found by us and others, both at low (Saintonge et al. 2011b) and high (e.g. Tacconi et al. 2013, Magdis et al. 2012b, G15, Scoville et al. 2017) redshift. Driven by the xCOLD GASS data at $\log(M_*/M_\odot)<10$ SFGs there is a tendency for the slope to steepen at the high mass tail, and requiring a second order fitting function, log $\mu_{gas}$ =0.35$_{0.15}$-0.085$_{0.05}$×(log($M_*$/M$_\odot$)-8.7$_1$). As in the discussion of the depletion time dependence on mass, conclusions on the distribution at low mass strongly depend on the metallicity correction of $\alpha_{CO}$ applied. This drop reflects the drop seen in the MS reference curve (W14, Schreiber et al. 2015), and is most likely *correlated with the quenching of galaxies beyond the Schechter mass* at all redshifts, the so-called 'mass quenching' and the formation of bulges (e.g. Peng et al. 2010). Finally the upper right panel of Figure 6 reflects the weak dependence $\mu_{gas}$ on size, as for $t_{depl}$, once one has marginalized over the other three parameters. Potential reasons for this lack of a size dependence are discussed in the previous section.

## 5. Summary and Discussion

We have updated and improved the analysis presented in G15 of the scaling relations of molecular depletion times and gas fractions with redshift and integrated galaxy parameters.

1) We have included new CO data emerging for *z*=0.5-2.5 MS-SFGs from the first three years of the PHIBSS2 survey on the IRAM NOEMA mm-interferometer and from ALMA (this paper, Freundlich et al. 2018, and DeCarli et al. 2016), as well as CO data for the low-mass extension of xCOLD GASS obtained with the IRAM 30m telescope by Saintonge et al. (2017). This increases the number of CO detections by 33% to 667 SFGs (compared to 500 in G15), and improves the statistical robustness at z=0.5-2.5 and the stellar mass parameter coverage.
2) We have added two independent, published analyses of the dust/gas content of *z*=0.4-4.5 SFGs obtained from fitting stacks of Herschel PACS/SPIRE SEDs (Santini et al. 2014, Béthermin et al. 2015). In comparison to the dust analysis presented in G15, which included fluxes from Magnelli et al. (2014) and analysis from Berta et al. (2016), the comparison of these three studies, based on similar data, gives an objective measure of the systematic uncertainties of the dust – SED methods and adds 135 further data points for a total of 647 measurements, 517 of which are at z≥0.4.
3) New data have emerged based on a single ~1mm broad-band emission observations (e.g. Scoville et al. 2016), and we have included those published data here, as well as 6 recent detections with NOEMA. Applying the prescriptions of Scoville et al. (2016), with the addition of a metallicity dependent gas to dust ratio, this adds 108 individual detections and 22 stacks of z=1-4.4 SFGs. Thus our overall data set contains 1444 measurements, about 41% more than in G15.
4) We have homogenized and brought onto the same calibration the parameters entering the analysis. For high-z we based stellar masses and star formation rates, wherever possible, on the 'ladder technique' of Wuyts et al. (2011a). The calibration of stellar masses and star formation rates of the *z*=0 galaxies in the xCOLD GASS survey has been recently updated by Saintonge et al. (2016, 2017) to be consistent with the Wuyts et al. (2011a) approach.
5) The most important new result of our study is the demonstration, shown in Figures 4-6, that the various methods and analyses presented here and elsewhere in the literature *converge to consistent quantitative scaling relations with redshift, specific star formation rate and stellar mass, if modest (0-0.21 dex) adjustments are made to the zero points of each of the techniques and data sets*. We note the agreement of our results and conclusions with the recent results of Scoville et al. (2017), obtained at the same time as our study and entirely with the 1mm single band dust technique in 708 galaxies. Where substantial differences have occurred in the recent literature, they can be arguably accounted for by one or several of the following:
   - different calibrations in the ancillary inputs of stellar mass, star formation rate and redshift,

- different assumptions about CO or dust mass conversion factors/functions, metallicity corrections, and main-sequence prescriptions,
- uncertainties in inferred parameter values and slopes, often driven by limited redshift coverage or, simply, by limited statistics, given the ±0.2 to ±0.28 dex errors of individual measurements of depletion time, star formation rate, etc.

The quantitative results for the fit parameters of equations (5) and (6) are summarized in Tables 3a and 3b and visualized in Figure 7, which shows the overall distributions of depletion time and $\mu_{gas}$ in the stellar mass- star formation rate plane, with the MS redshift dependence removed (eq.(1)). The key findings shown in Figure 7, in excellent agreement with and improve the precision of the earlier results presented in Table 1, are:

1. The depletion time scale, or the gas regulator's efficiency, drops relatively slowly with cosmological epoch ($t_{depl}$~$(1+z)^{-0.6}$), about a factor of 2 between $z$=0 and 2.5. This suggests that MS star formation is driven by similar physical processes at high- and low-redshift. The possible downturn of the depletion time scale (and also gas fraction, seen in the residual plots in the bottom row of Figure 7) at the low-mass tail could be due to the very sensitive to the metallicity dependence of $\alpha_{CO}$, as discussed in section 4.2.3. We refer to section 4.3 of G15 for the discussion of possible origins the shallow dependence on redshift, which appears to be shallower than $H(z)^{-1}$, and could suggest that depletion time scale and star formation efficiency are set by local processes within molecular clouds. As a result galactic molecular gas reservoirs track the cosmological evolution of the star formation rate, which in turn tracks the evolution of baryonic accretion into galaxies.
2. As one steps along in *sSFR* at fixed *z* and $M_*$, gas fractions and star formation efficiency (log($1/t_{depl}$)~$\log\mu_{gas}$~$0.5\times\log\delta MS$ ) increase, in approximately equal measure (Top row of Figure 7). The former is a measure of the value of the gas accretion rate at the observed redshift, and is modulated by mergers and variations in gas transport along cosmic web filaments. The latter is probably a measure of internal galaxy properties and/or bulge/disk ratio, and may be a proxy for gas density.
3. As one steps along the main-sequence, from lower to higher stellar mass, the star formation efficiency stays roughly constant, while the gas fractions drop at high stellar mass, as does the *sSFR*. As proposed by others (W14, Schreiber et al. 2015), this drop is probably strongly correlated with the process of internal 'mass' quenching (Peng et al. 2010).
4. Our preliminary finding is that the residual depletion time, after removing all other parameter dependencies, does not significantly depend on galaxy size, and thus not on surface density, above and beyond what may already be encapsulated in the galaxy integrated *sSFR*. However, the sizes entering this analysis come from rest-frame optical stellar light. In highly dusty systems, the stellar and gas distributions may have significantly different radial scales. Spatially resolved measurements of the molecular gas and dust distributions are urgently needed for a more robust test of the size/surface density dependence.
5. The quality and the remarkable congruence of data obtained with different methods and calibrations have allowed us to test for second order effects, beyond the simple single power laws in *(1+z), sSFR* and $M_*$. We find significant evidence for curvature in the $\mu_{gas}$-z relation, which tracks the *sSFR*-evolution, and perhaps indicates deviations from a simple ideal gas regulator with depletion time as its fastest time clock (Lilly et al. 2013). The mass dependencies of $t_{depl}$ and $\mu_{gas}$ may also require a curvature term at low stellar masses, although measurements in this regime of sub-solar metallicity gas, are challenged by our relatively poor knowledge of the metallicity dependence of the CO conversion factor and the dust to gas ratio. None of the other scaling relations require significant deviations from simple power-laws. In the stellar mass-star formation rate plane $t_{depl}$ only shows a vertical variation, while $\mu_{gas}$ varies in both coordinates (Figure 7 left top and bottom).

The scaling relations presented in this paper improve on previous work, since they are based on larger data sets available from our own work and from the literature, and they are derived using consistent assumptions for *SFR*s, stellar masses, and molecular gas and dust conversions. With these relations, it is now possible to determine molecular gas masses and depletion time scales with an accuracy and scatter of ±0.1 dex or better in relative terms and in sample averages, and ±0.2 to ±0.25 dex of individual galaxies, including systematic uncertainties.

***Acknowledgements***

*We are grateful to the staff of the IRAM facilities for the continuing excellent support of the large CO survey programs we have been analyzing in this paper. We also thank Albrecht Poglitsch and Matt Griffin, the PACS and SPIRE teams and ESA for their excellent work on the Herschel instruments and mission. Finally, we thank the referee, whose suggestions have led to a clearer manuscript overall. This paper makes use of the following ALMA data: ADS/JAO.ALMA#2013.1.00092.S. ALMA is a partnership of ESO (representing its member states), NSF (USA) and NINS (Japan), together with NRC (Canada), MOST and ASIAA (Taiwan), and KASI (Republic of Korea), in cooperation with the Republic of Chile. The Joint ALMA Observatory is operated by ESO, AUI/NRAO and NAOJ.*

## Appendix A: Accuracy of Inferred Parameters

In this section we briefly investigate how accurately the various model parameters discussed in this paper can be inferred from the current datasets. The goal is to assess the impact of limited redshift coverage and limited source numbers per parameter bin. We run models with 2 or more redshift bins, limited mass and sSFR coverage, and with varying numbers of sources. To do the exercise we create model data sets from Monte-Carlo realizations with Gaussian uncertainties driven by the actual data, and fit these in the same way as the data discussed earlier in the paper. We create multiple realizations of such data sets and then infer the model parameter distributions, relative to the input parameters.

As elsewhere in this paper we assume that the model-data can be described by products of power-laws, transforming in log-log space into sums of log-linear functions:

$$\begin{aligned}\log y = A_{in} &+ B_{in} \times \log(1+z) + G_{in} \times \log^2(1+z) \\ &+ C_{in} \times \log(\delta MS) \\ &+ D_{in} \times \log(\delta M) + H_{in} \times \log^2(\delta M) \quad \text{(A1).}\end{aligned}$$

The symbols are the same as in equations (5) and (6), where the dependent variable is $y$=$t_{depl}$(Gyr) or $y$=$\mu_{gas}$=$M_{molgas}/M_*$. To construct the model data sets, we diced values of $z$, $logM_*$ and $log\delta MS$ in bins centered on mean values listed in the second, fourth and fifth column of Table A1, over an interval of twice the values listed as sub-scripts in the same columns. We then computed the intrinsic value of *log y* from equation (A1) and diced the mock values of *log y*, $logM_*$ and $log\delta MS$ by assuming a scatter of 0.24 (0.32), 0.13 (0.18) and 0.22 (0.25) dex on the MS (δMS<0.4) (and in parentheses above the MS, $\delta MS$>0.4). We then solved for the best-fit parameters A, B, C, etc. with a classical leveraged, weighted multiple linear regression model, based on minimizing $\chi^2$. Note that in those cases where we tested for the detectability of curvature (square terms) in log($1+z$) or log($\delta M_*$), we solved for the parameter of square-term directly (G or H) instead of the formulation in equation (6).

Columns 7-12 of Table A1 give the resulting means and their 1σ dispersions (subscripts) of the output parameter distributions, after repeating this exercise multiple times. As an example Figure A1 shows the derived parameter distributions (relative to the input values of the model) for the first case in Table A1, which consisted of multiple Monte-Carlo realizations of two sub-samples at $z$~1.2 and 2.3, with 20 galaxies each. While the mean values of these distributions are no more biased than expected from their fit uncertainties, the dispersions of the parameter distributions exhibit substantial differences. These parameter distributions are generally symmetric around a mean (with some offset from the input value) but have wings that can substantially deviate from Gaussian shape (e.g. the distribution of B).

The examples shown in Table A1 demonstrate that the MS-offset and stellar mass dependencies can be inferred with modest bias and a dispersion of ±0.05 to ±0.2 dex from data sets with N~40 sources in one or two redshift slices, as long as the coverage in $\delta MS$ and $\delta M_*$ is >1dex. Uncertainties grow rapidly once the coverage in these parameters is smaller, or is biased significantly toward above-MS values**,** as has been the case in many of the studies in the literature up until now (see section 2.4).

The simulations show that accurate (≤0.15dex) determinations of the redshift dependence of $t_{depl}$ or $\mu_{gas}$ cannot be done from modest data sets based only on two redshift slices but require a wider range of z, including coverage at $z$~0, such as the dataset compilations that we have used in this paper. This constraint is especially relevant for determination of gas fractions, since $log\mu_{gas}(log(1+z))$ exhibits curvature (Figure 6), and the slope $|dlog\mu/dlog(1+z)|$ increases toward smaller z. The last three simulations in Table A1, with 40 galaxies in two redshift slices, with the two slices increasing from 0.02/0.5 to 1.2/23 and 2.3/3.4, clearly show how this curvature term, if not included, leads to systematically decreasing slopes $dlog\mu/dlog(1+z)$=3.6, 1.8 and 0.9 for the three combinations above. This offers a straightforward explanation of why the FIR and 1mm dust data sets, which range from z~0.4 to 4, yield systematically shallower redshift dependences of $\mu_{gas}(z)$ than the CO data (Tables 1, 3a and 3b), and also why Scoville et al. (2016, 2017) find shallower slopes than our best values in Tables 3a and 3b.

A determination of all parameters to better than ±0.1 dex requires several hundred galaxies covering a wide range of the parameter space in $z$, $\delta MS$ and $\delta M_*$. With such more extensive data sets, as presented in this paper, it is then possible to detect with significance curvature in the relations (such as for $log\mu_{gas}(log(1+z))$ or $log\mu(\delta M_*)$ in Figure 5). Finally, the parameter uncertainties estimated from the simple models considered here are similar to those obtained in the actual data (Tables 3a and 3b), indicating that there are no major hidden parameter dependencies or large undetected systematic errors.

# Figures

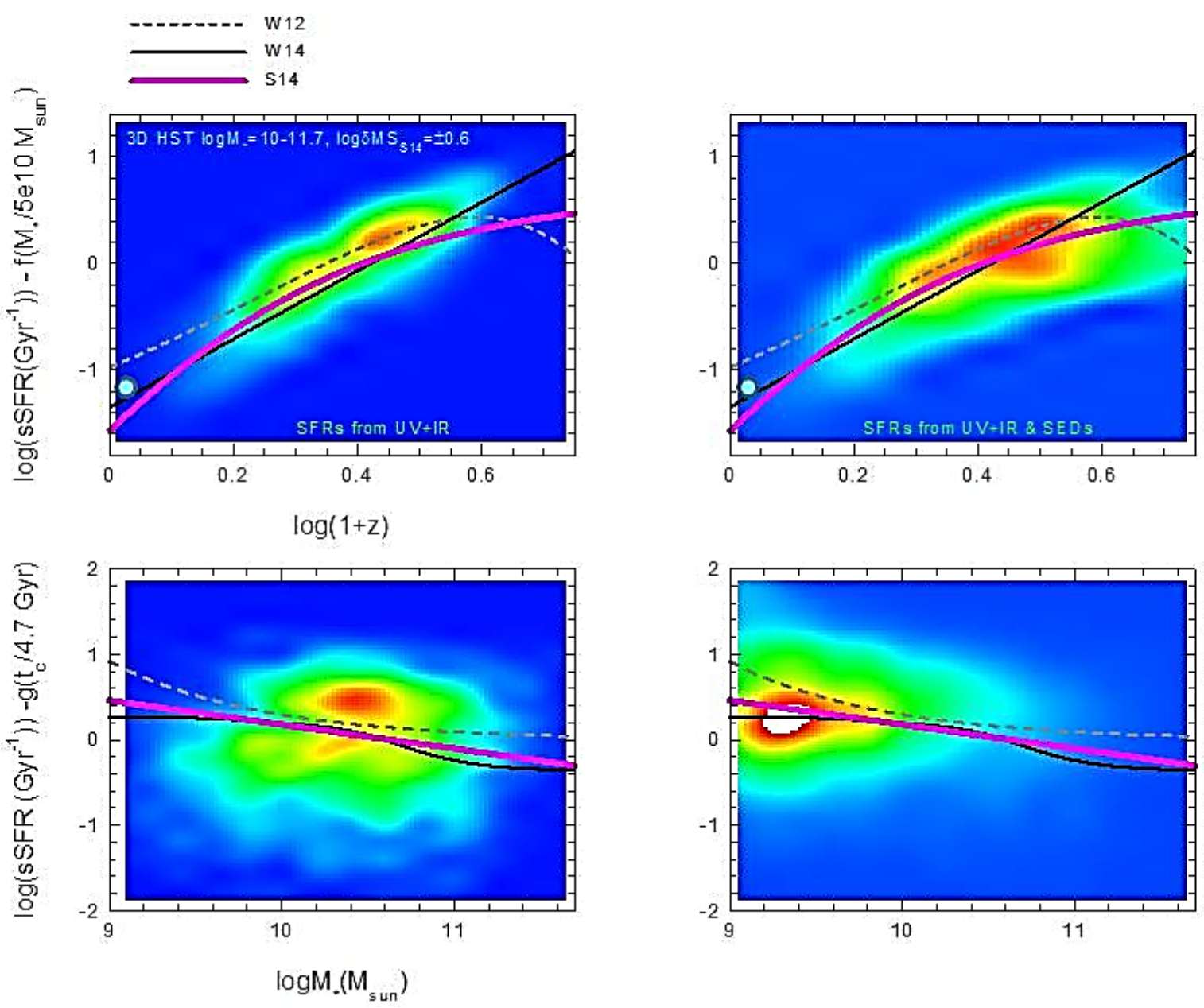

**Figure 1.** Specific star formation rates, *sSFR=SFR/M∗* ($Gyr^{-1}$) as a function of *log (1+z)* at *log(M∗/*$M_\odot$*)*=10.7 (top panels) and as a function of log*M∗* (bottom panels). The color distributions represent the distribution of galaxies in the 3D-HST survey on a linear scaling (Brammer et al. 2012, Skelton et al. 2014, Momcheva et al. 2016). In the left panels we show 3D-HST galaxies with *log(M∗/*$M_\odot$*)*=10-11.7, log*δMS*=±0.6, which have individual 24μm Spitzer, or 70 μm, 100μm or 160μm Herschel detections, so that a IR+UV luminosity can be computed (Wuyts et al. 2011a). The right panels in addition include galaxies between *log(M∗/*$M_\odot$*)*=9-10 and galaxies across the entire mass range with only an SED-based *SFR*, typically resulting in underestimated *SFRs*, which is particularly relevant at high-z, low *log(M∗)* and below the MS. We used the S14 MS prescription (Equation (1) in the main text) to correct all galaxies to the same mass of *log(M∗/*$M_\odot$*)*=10.7 in the left plot, and to the same redshift (*z*=1.5, $t_c$=4.7 Gyr) in the right plot. The solid magenta, dotted grey and solid black lines denote the S14, W12 and W14 prescriptions of the MS, respectively. The cyan circle in the top panels denotes the location of the MS-line for the SDSS sample. It is clear that all three prescriptions have their advantages and their disadvantages. For this paper we use S14 as our default prescription (Equation (1)).

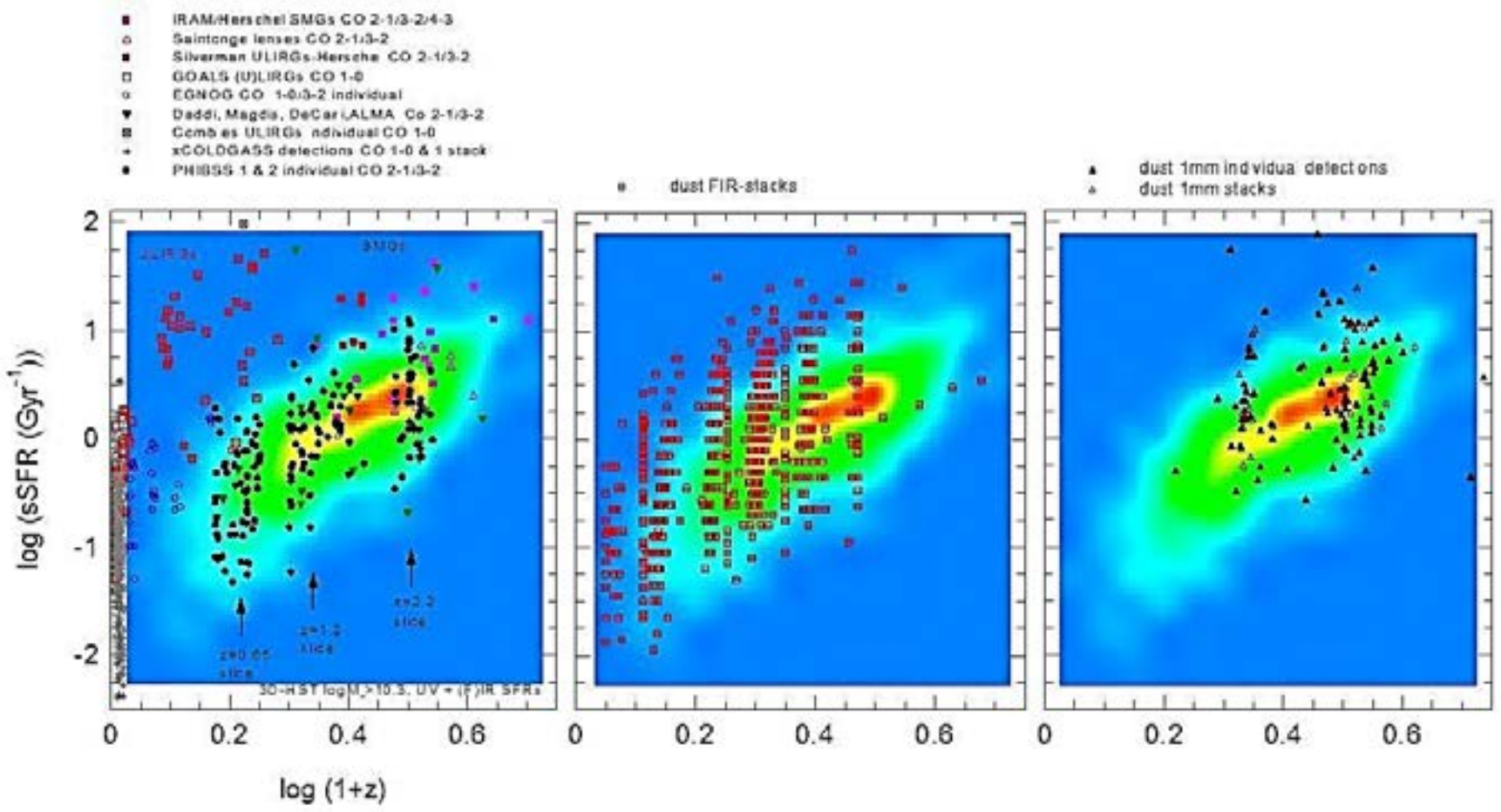

**Figure 2.** Distribution of the SFGs in our samples in the *logsSFR-log(1+z)* plane, superposed on the distribution of the 3D-HST parent sample (color). The various CO data sets (all individual detections at >3σ) are shown in the left panel. The stacks of FIR dust SED data sets are shown in the middle panel (≥4 σ detections), and the 1mm dust photometry points are shown in the right panel, with filled symbols denoting individual detections (≥3.8σ) and the open symbols denoting stacks. The various symbols are explained above the figure (see section 3).

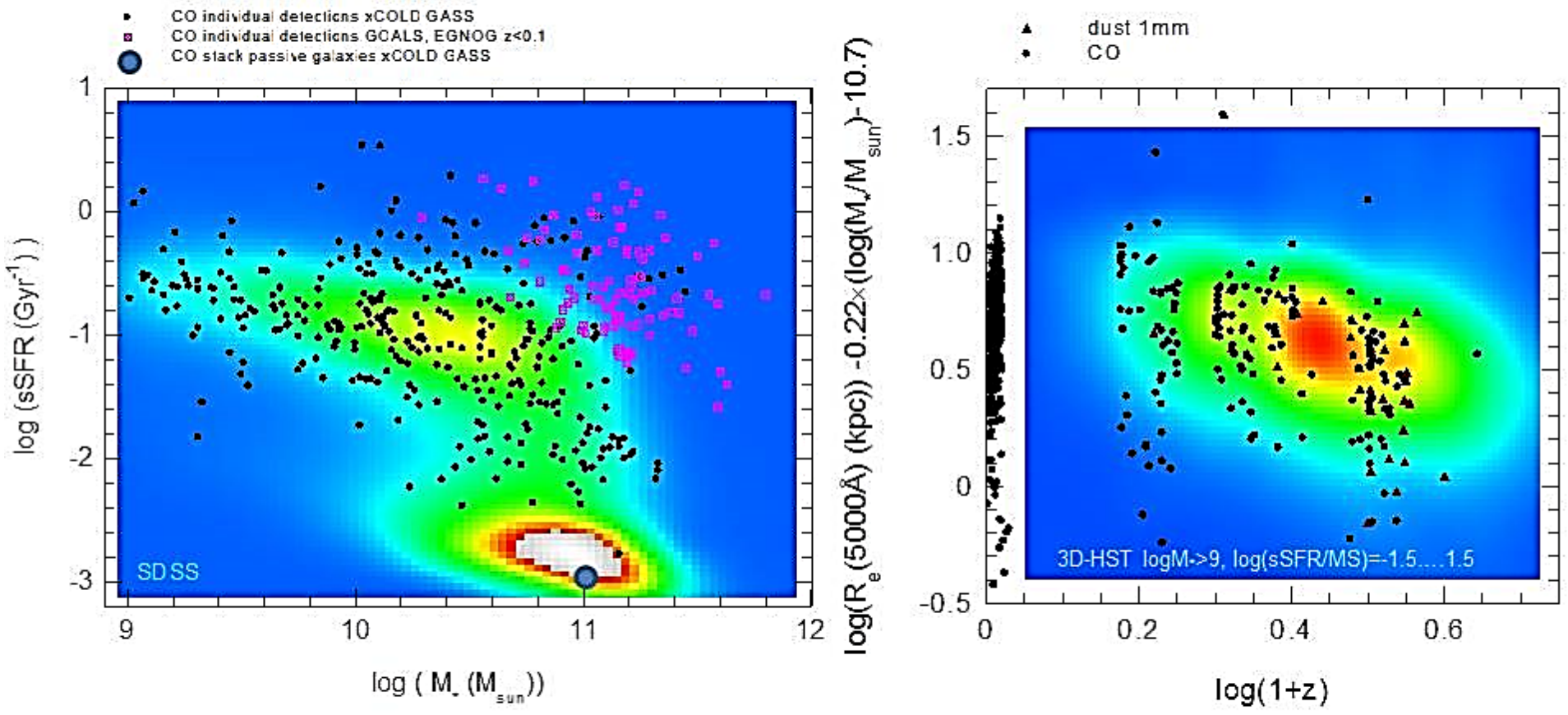

**Figure 3.** Distribution of the SFGs in our samples in the *logsSFR-log(M∗)* plane (left), and the *logR*$_e$*-log(1+z)* plane (right). In the left panel our CO data from the xCOLD GASS sample (filled black circles) and the GOALS sample (open crossed magenta squares), including a stack of below-MS xCOLD GASS galaxies (large blue circle) are superposed on the distribution of the SDSS parent sample (color with a linear scale). For the distribution of sizes in the right panel we mainly used effective H-band radii from 3D-HST data and converted these sizes to 5000 Å effective radii using the van der Wel et al. (2014) prescriptions. The data are superposed on the distribution of $\log(M_*/M_\odot)\geq 9$, $\log\delta MS=\pm 0.6$ 3D HST sample (see text in section 3).

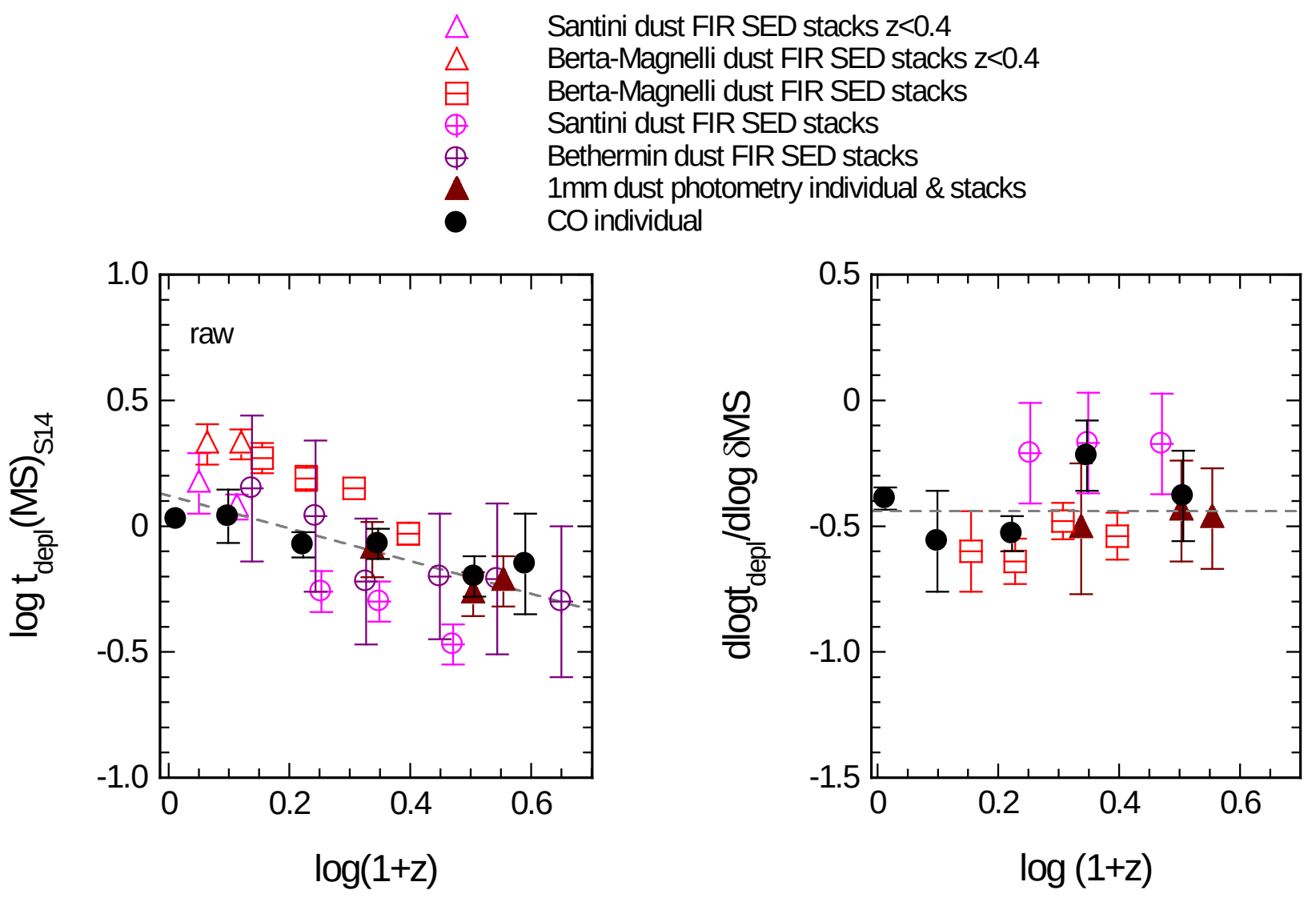

**Figure 4.** Redshift dependence of MS molecular depletion time $t_{depl}(MS)$ (left) and slope $C_t=dlogt_{depl}/dlog\delta MS$ (right), when the 'raw' zero points of the original papers are used. Left panel: depletion time at the main-sequence reference line ($\delta MS$=1, for $C_t$= -0.44) in bins of z, as a function of *log(1+z)*, for different methods (CO: filled black circles, dust-FIR/submm SED: open red rectangles with horizontal bar (Magnelli et al. 2014, G15, Berta et al. 2016), open green circles (Béthermin et al. 2015), open magenta crossed circles (Santini et al. 2014), dust-1mm-photometry: brown filled triangles (Scoville et al. 2014, 2016, DeCarli et al. 2016, Barro et al. 2016, Tadaki et al. 2017, Lilly et al.in preparation, Dunlop et al. 2017, this paper). Open red and magenta triangles mark the z<0.4 dust FIR SED results, which may be affected by dust in the atomic HI ISM and were excluded from most of the further analysis. The dotted grey line is the unweighted fit: $A_t$=0.089, $B_t$=-0.62. Right panel: slope $C_t$ of depletion time in $logt_{depl}$-$log(\delta MS)$ plane (equation (5)) as a function of *log(1+z)*. Error bars are ±1σ and the dotted grey line ($C_t$=-0.44) is the best fit global fit value for all data.

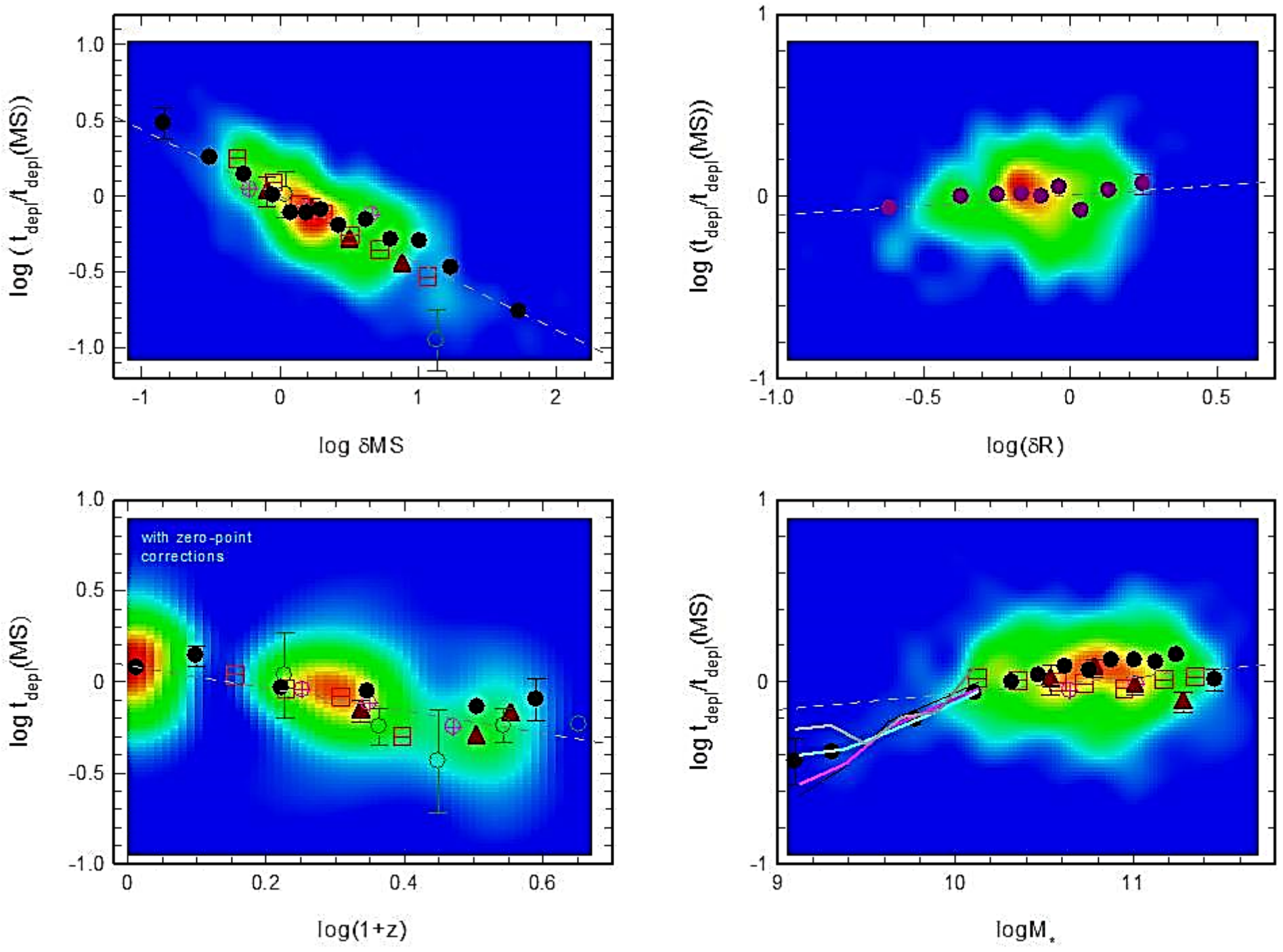

**Figure 5.** Dependence of molecular depletion time, $t_{depl}=M_{molgas}/SFR$ on *z*, *sSFR*, $M_*$ and $R_e$, after we introduce zero-point corrections for the different methods to minimize the scatter in $t_{depl}$(MS)-(*1+z*) relation (bottom left: +0.02 dex for CO, -0.22 dex for Magnelli/Genzel/Berta, -0.02 dex for Béthermin, +0.21 dex for Santini, -0.003 dex for Scoville 1mm. With these zero point corrections, the 1σ scatter around the best fitting slope (unweighted: $B_t$=-0.62) decreases from ±0.15 dex in the left panel of Figure 4 to ±0.066 dex. The colored distribution marks the overall distribution of our data, and the large symbols the binned averages (with the same nomenclature as in Figure 4), all now zero-point adjusted. The dotted grey line is the global fit: A=0.09, B=-0.62. Top left: after removal of the redshift dependence (from above) this panel displays the dependence of the depletion time residuals perpendicular to the reference main-sequence line, for all data (colored distribution) and for the different binned averages separately, as a function of log *δMS*, again with the same symbols as in Figure 4. The dotted grey line is the best fit global fit ($C_t$=-0.44). Bottom right: dependence of the depletion time residuals on the MS-reference line (*δMS*=1) along the main-sequence (= as a function of $logM_*$), after removal of the best fit, redshift trend ($A_t$=0.09, $B_t$=-0.62, $C_t$=-0.44). The dotted grey line is the best global fit ($D_t$=0.09). Symbols are the same as in the other panels. The various continuous lines on the left indicate the data trends at low $M_*$, if in the metallicity correction of $\alpha_{CO}$ instead of equation (2), no (black), equation 31 of Bolatto et al. (2013, magenta), equation 7 of G15 (see Genzel et al. 2012, cyan), or equation 24 of Accurso et al. (2016) are chosen. Top right panel: Depletion time scale residuals as a function of the rest-frame optical (5000 Å) effective radius for a Sersic model, normalized by the average size of the star forming galaxy population at that redshift and stellar mass (van der Wel et al. 2014), after removing the redshift, *sSFR* and mass dependences ($A_t$=0.09, $B_t$=-0.62, $C_t$=-0.44, $D_t$=0.1). The dotted grey line is the best global fit ($E_t$=0.11). Symbols are the same as in the other panels. All dashed best fit lines come from a global, multi-parameter fit with equation (5), weighted by the inverse squares of the uncertainty of each data point.

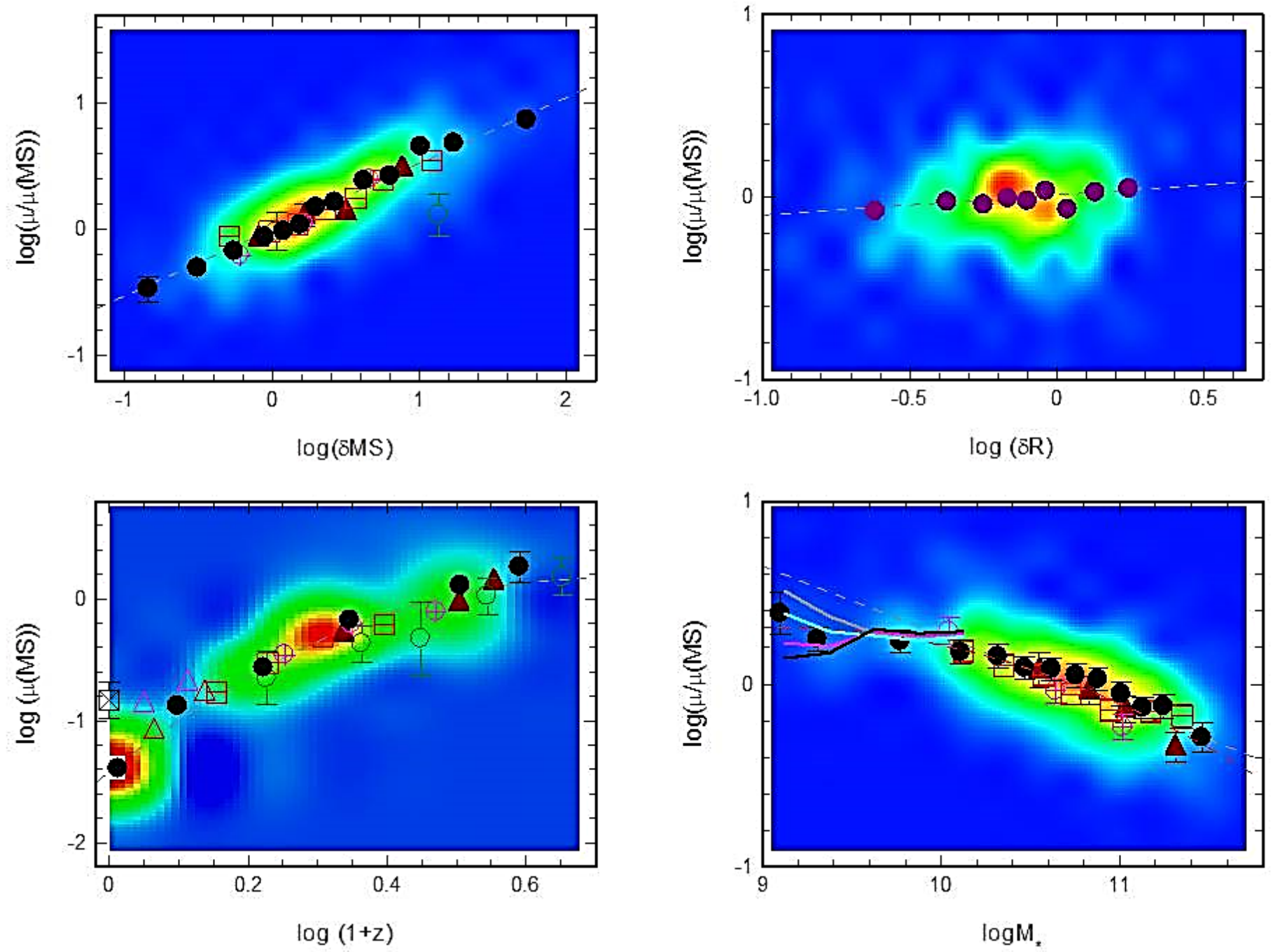

**Figure 6.** Scaling relations of $\mu_{gas}=M_{molgas}/M_*$ with redshift (bottom left), specific star formation rate offset $\delta MS$ (top left), stellar mass (bottom right), and normalized optical radius (top right), for the binned data sets (same nomenclature and analysis procedure as in Figure 5) and the individual data points (colored distributions). The open crossed black square denotes the total gas fraction (HI+H2), obtained in the COLD GASS survey at $log(M_*/M_\odot)=10.7$ (Saintonge et al. 2011a). All dashed best fit lines come from a global multi-parameter fit with equation (6), weighted by the inverse squares of the uncertainty of each data point. The various continuous lines in the bottom right panel indicate the data trends at low $M_*$, if different metallicity corrections of $\alpha_{CO}$ are chosen instead of equation (2): no correction (black), equation 31 of Bolatto et al. (2013, magenta), equation 7 of Genzel et al. (2015, see Genzel et al. 2012, cyan), equation 24 of Accurso et al. (2016).

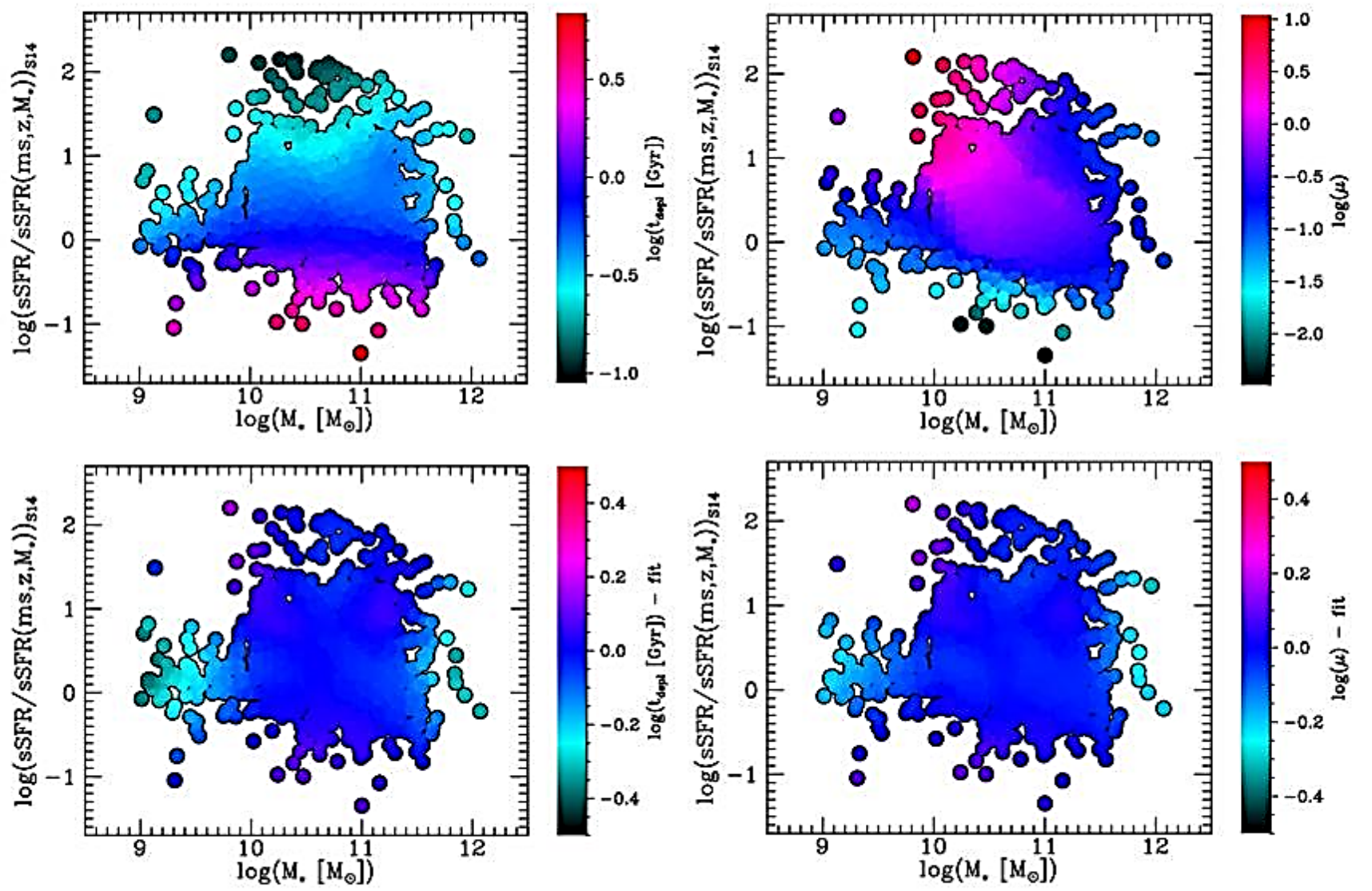

**Figure 7.** Graphical summary of the two-dimensional distributions of depletion time scale $t_{depl}$ (left) and molecular gas to stellar mass ratio, $\mu_{gas}$(right) in the stellar mass – specific star formation rate plane, after removing the redshift dependencies. The top row shows the smoothed data (combining CO and dust techniques), while the bottom row gives the residuals between the data and the best fitting scaling relations (Table 3). The only remaining features in the bottom plots is the possible downturn of the depletion time scale and gas fraction at the low-mass tail, which is very sensitive to the metallicity dependence of $\alpha_{CO}$, as discussed in section 4.2.3. In all panels, the color-coding corresponds to LOESS-smoothed quantities. The LOESS method recovers underlying mean values by accounting for neighboring data points (and errors) is essentially a running local average (Cappellari et al. 2013).

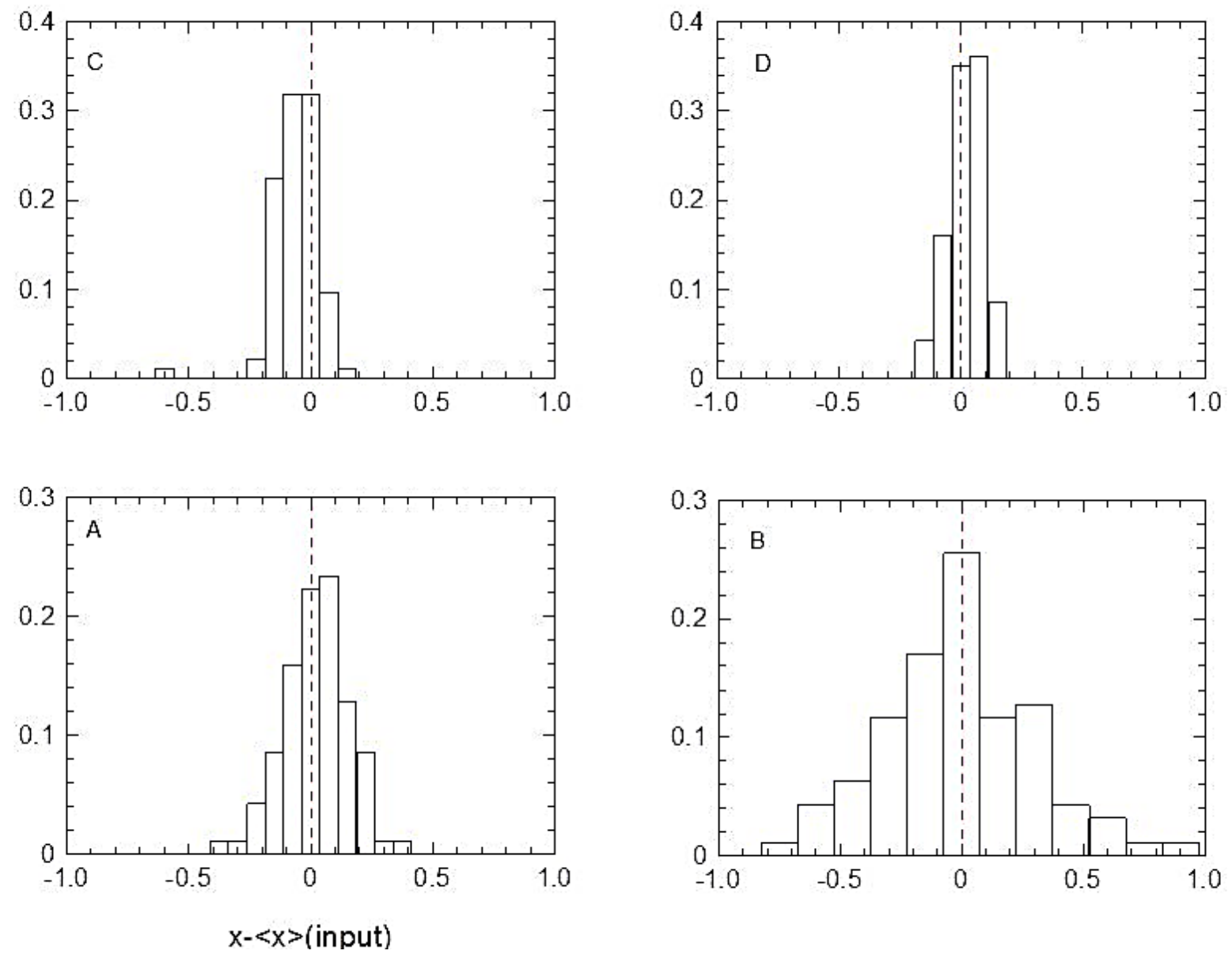

**Figure A1.** Distribution of fit parameters $A_t$, $B_t$, $C_t$ and $D_t$, all relative to the input model values, for the first simulation in Table A1, made from multiple Monte-Carlo realizations of two sub-samples at $z$~1.2 and 2.3, with 20 galaxies each. While the mean values of these distributions are no more biased than expected from their fit uncertainties, the dispersions of the parameter distributions exhibit substantial differences. While the mass- and MS-offset dependencies can be reasonably well determined from a N=40 sample with ±0.65 and ±0.8 dex coverage in these parameters, two redshift slices are not sufficient to determine the redshift dependence of $t_{depl}$ to  better than a factor of 2-3 (see text in Appendix A1.

**Table 1. Summary of scaling relations in the current literature**

| Method | z-range | $dlogt_{depl}/dlog(1+z)$ | $dlog\mu_{gas}/dlog(1+z)$ | $dlogt_{depl}/dlog(\delta MS)$ | $dlogt_{depl}/dlog(M_*)$ | $dlog\mu_{gas}/dlog(M_*)$ | scaling with other parameters | $dSFR/dM_{molgas}$ $d\Sigma_{SFR}/d\Sigma_{molgas}$ | Reference |
|---|---|---|---|---|---|---|---|---|---|
| CO COLD GASS galaxy integrated | 0.025-0.05 (N=440 incl. non-detects) | | | $-0.5_{0.1}$ | $+0.36_{0.1}$ | $-0.24_{0.1}$ | $dlogt_{depl}/dlog\Sigma_*\sim -0.5$, $dlogt_{depl}/dlog\Sigma_{SFR}\sim -0.36$ no bars/interactions, no or weak AGN | $1.18_{0.24}$ | Saintonge+11a,b,12, 13,16a, Huang & Kauffmann 14, 15 Accurso+16 |
| CO HERACLES galaxy 1 kpc resolved | 6-35e-4 (N=30) | | | $-0.25_{0.25}$ | $+0.2_{0.15}$ | | weak with radius, spiral arms, surface density, except nucleus | $1_{0.15}$ (>1.5 kpc) | Bigiel +08,11 Leroy+08,13, Schruba +11 |
| CO PHIBSS1 | 1-2.5 (N=52 detects.) | $-0.7_{0.3}$ | $+2.7_{0.4}$ | $-0.32_{0.2}$ | | $-0.6_{0.15}$ | | $1.1_{0.2}$ | Tacconi+10,13 Genzel+10 |
| CO + dust FIR/submm SED (plus literature) | 0-4 (N=131 CO detects. 15 FIR stacks) | $-0.8_{0.3}$ | $+2.6_{0.4}$ | $-0.2_{0.1}$ | | | 'two mode' SF, with strong increase of bursts for δMS>4 | $1.25_{0.15}$ | Daddi +10b, Magdis+12a, Sargent+14, Bethermin+15 |
| dust FIR SED | 0.3-2 (N=121 stacks) | $-1.5_{0.4}$ | $+1.6_{0.5}$ | $-0.3_{0.1}$ | | | | | Santini+14 |
| CO PHIBSS 1+2 +dust FIR SED (plus literature) | 0-2.5 (N=500 CO >4σ detects., 512 FIR stacks) | $-0.34_{0.3}$ | $+2.7_{0.2}$ | $-0.49_{0.05}$ | $+0.01_{0.1}$ | $-0.37_{0.1}$ | | $1.1_{0.15}$ increases with δMS | Genzel+15 (including data from Magnelli +14, cf. Berta+16) |
| dust 1mm | 1-4.4 (N=51 >4σ detects., 35 stacks) | $-1.2_{0.4}$ | $+1.8_{0.5}$ | $-0.55_{0.1}$ | $+0.23_{0.2}$ | $-0.02_{0.2}$ | | $0.9_{0.1}$ | Scoville+16 |
| dust 1mm | 0.3-4.5 (N=708) | $-1.05_{0.05}$ | $+1.8_{0.14}$ | $-0.7_{0.02}$ | $-0.01_{0.01}$ | $-0.7_{0.04}$ | | 1.0 | Scoville +17 |
| semi-analytic model | All | | $+2.5_{0.3}$ | | | | | | Popping+15 |
| Illustris hydro sim | All | | $+1.8_{0.4}$ | | | $-0.6_{0.2}$ | | | Genel+14 |
| Eagle hydro sim | All | $-1.3_{0.4}$ | $+1.5_{0.4}$ | | | | | | Lagos+15a |

Note: the subscripts in columns 3-9 are the formal uncertainties as listed in the individual studies.

**Table 2. Source Physical Properties**

(Full version available online)

| Source | Survey/Method | Redshift | log $M_{gas}$ ($M_\odot$) | log $M_*$ ($M_\odot$) | log SFR ($M_\odot$ $yr^{-1}$) | Log $r_{eff}$ (kpc) |
|---|---|---|---|---|---|---|
| G3-7652 | PHIBSS2/CO | 0.502 | 9.8 | 10.5 | 0.9 | 0.90 |
| 838945 | PHIBSS2/CO | 0.502 | 9.9 | 10.7 | 0.6 | 1.03 |
| 834187 | PHIBSS2/CO | 0.502 | 10.5 | 11.1 | 1.3 | 0.84 |
| 831870 | PHIBSS2/CO | 0.502 | 10.2 | 10.2 | 1.5 | 0.86 |
| GN3-5128 | PHIBSS2/CO | 0.503 | 9.9 | 10.3 | 0.7 | 0.81 |
| G3-4097 | PHIBSS2/CO | 0.509 | 10.3 | 11.3 | 1.6 | 1.17 |
| G3-8310 | PHIBSS2/CO | 0.509 | 10.1 | 10.4 | 1.0 | 0.91 |
| … | … | … | … | … | … | … |
| stack Magnelli+14, Berta+16 | PEP+HerMES/dust | 0.400 | 10.6 | 11.2 | 0.0 | |
| stack Magnelli+14, Berta+16 | PEP+HerMES/dust | 0.401 | 10.5 | 10.9 | 1.4 | |
| stack Magnelli+14, Berta+16 | PEP+HerMES/dust | 0.401 | 10.2 | 10.6 | 1.3 | |
| stack Magnelli+14, Berta+16 | PEP+HerMES/dust | 0.405 | 9.9 | 10.1 | 1.1 | |
| stack Magnelli+14, Berta+16 | PEP+HerMES/dust | 0.406 | 10.5 | 11.2 | 1.7 | |
| stack Magnelli+14, Berta+16 | PEP+HerMES/dust | 0.407 | 11.1 | 11.6 | 1.6 | |
| stack Magnelli+14, Berta+16 | PEP+HerMES/dust | 0.407 | 10.0 | 10.4 | 1.3 | |
| … | … | … | … | … | … | … |

**Table 3a. Fit parameters for equations (5) obtained from error-weighted, multi-parameter regression:**

$$\log t_{depl}\,(\mathrm{Gyr}) = A_t + B_t * \log(1+z) + C_t * \log(sSFR/sSFR(MS,z,M_*)) + D_t * (\log M_* - 10.7) + E_t * \log(R_e/R_{e0}(z,M_*))$$

| Data | parameter | N | $\chi^2_r$ | $A_t$ [a] | $B_t$ | $C_t$ | $D_t$ | $E_t$ |
|---|---|---|---|---|---|---|---|---|
| all (error weighted) (after removal of z<0.4 dust points) | $t_{depl}$ (Gyr) S14 | 1309 | 0.86 | $+0.09_{0.03}$ | $-0.62_{0.08}$ | $-0.44_{0.03}$ | $+0.095_{0.03}$ | - |
| all (err.w.) including z=0.1-0.4 dust points | $t_{depl}$ (Gyr) S14 | 1444 | 0.85 | $+0.12_{0.03}$ | $-0.66_{0.1}$ | $-0.45_{0.03}$ | $+0.09_{0.03}$ | - |
| SFGs with 5000Å $R_e$ (error weighted) | $t_{depl}$ (Gyr) S14 | 512 | 0.85 | $+0.006_{0.02}$ | | | | $+0.11_{0.08}$ |
| all (equal weight per z-bin) | $t_{depl}$ (Gyr) S14 | 1309 | 0.73 | $+0.09_{0.03}$ | $-0.52_{0.1}$ | $-0.45_{0.03}$ | $+0.02_{0.04}$ | - |
| CO (error w.) | $t_{depl}$ (Gyr) S14 | 667 | 0.89 | $+0.06_{0.03}$ | $-0.44_{0.13}$ | $-0.43_{0.03}$ | $+0.17_{0.04}$ | - |
| dust-FIR (error w.) | $t_{depl}$ (Gyr) S14 | 512 | 0.56 | $+0.25_{0.13}$ | $-1.0_{0.5}$ | $-0.53_{0.07}$ | $-0.07_{0.07}$ | - |
| dust-1mm (error w.) | $t_{depl}$ (Gyr) S14 | 130 | 1.3 | $+0.42_{0.45}$ | $-0.7_{0.9}$ | $-0.64_{0.15}$ | $-0.27_{0.2}$ | - |
| **best (with bootstrap errors)** | $t_{depl}$ (Gyr) S14 | 1309 | | $\mathbf{+0.09_{0.05}}$ | $\mathbf{-0.62_{0.13}}$ | $\mathbf{-0.44_{0.04}}$ | $\mathbf{+0.09_{0.05}}$ | $\mathbf{+0.11_{0.12}}$ |
| **best (with bootstrap errors)** | $t_{depl}$ (Gyr) W14 | 1309 | | $\mathbf{+0.002_{0.04}}$ | $\mathbf{-0.37_{0.08}}$ | $\mathbf{-0.40_{0.04}}$ | $\mathbf{+0.17_{0.05}}$ | - |
| all (error weighted) (after removal of z<0.4 dust points) | $t_{depl}$ (Gyr) S14 fit *sSFR* instead of $\delta MS$ | 1309 | 0.86 | $-0.53_{0.04}$ | $0.95_{0.15}$ | $-0.45_{0.04}$ | $-0.08_{0.05}$ | |

**Table 3b. Fit parameters for equations (6) obtained from error-weighted, multi-parameter regression:**

$$\log M_{molgas}/M_* = A_\mu + B_\mu*(log(1+z)\text{-}F_\mu)^\beta + C_\mu*log(sSFR/sSFR(MS,z,M_*)) + D_\mu*(logM_*\text{-}10.7) + E_\mu*log(R_e/R_{e0}(z,M_*))^c$$

| Data | parameter | N | $\chi^2_r$ | $A_\mu$ [a] | $B_\mu$ | $F_\mu$ [b] | $C_\mu$ | $D_\mu$ | $E_\mu$ |
|---|---|---|---|---|---|---|---|---|---|
| all (error w.)<br>β=2 in log(1+z)<br>β=0 | $\mu$=$M_{molgas}/M_*$<br>S14 | 1309 | 0.7<br><br>0.95 | $+0.07_{0.15}$<br><br>$-1.35_{0.03}$ | $-3.8_{0.4}$<br><br>$+2.95_{0.08}$ | $+0.63_{0.1}$<br><br>- | $+0.53_{0.03}$<br><br>$+0.54_{0.03}$ | $-0.33_{0.03}$<br><br>$-0.31_{0.03}$ | - |
| all with optical $R_e$<br>(error w.) | $M_{molgas}/M_*$<br>*S14* | 512 | 0.85 | $-0.012_{0.03}$ | - | - | - | - | $+0.11_{0.07}$ |
| all (equal weight<br>per z-bin) | $M_{molgas}/M_*$<br>*S14* | 1309 | 0.7 | $+0.17_{0.15}$ | $-3.25_{0.4}$ | $+0.7_{0.1}$ | $+0.53_{0.03}$ | $-0.38_{0.04}$ | - |
| CO (ew.) β=2<br>β=0 | $M_{molgas}/M_*$<br>*S14* | 667 | 0.8<br>0.89 | $+0.19_{0.24}$<br>$-1.4_{0.03}$ | $-3.38_{0.8}$<br>$+3.1_{0.12}$ | $+0.7_{0.17}$<br>- | $+0.56_{0.03}$<br>$+0.56_{0.03}$ | $-0.30_{0.04}$<br>$-0.28_{0.04}$ | - |
| dust-FIR (error w.)<br>β=0 | $M_{molgas}/M_*$<br>*S14* | 512 | 0.6 | $-0.9_{0.1}$ | $+1.9_{0.3}$ | - | $+0.46_{0.15}$ | $-0.40_{0.07}$ | - |
| dust-1 mm (error w.)<br>β=0 | $M_{molgas}/M_*$<br>*S14* | 130 | 1.2 | $-0.44_{0.5}$ | $+1.0_{0.8}$ | - | $+0.36_{0.15}$ | $-0.52_{0.2}$ | - |
| all (error w.)<br>β=2 in log(1+z)<br>β=0 | $M_{molgas}/M_*$<br>*S14* | 1309 | 0.7<br><br>0.95 | $+0.07_{0.15}$<br><br>$-1.35_{0.03}$ | $-3.8_{0.4}$<br><br>$+2.95_{0.08}$ | $+0.63_{0.1}$<br><br>- | $+0.53_{0.03}$<br><br>$+0.54_{0.03}$ | $-0.33_{0.03}$<br><br>$-0.31_{0.03}$ | - |
| **best**[b] β=2<br>β=0 | $M_{molgas}/M_*$<br>*S14* | 1309 | | $\mathbf{+0.12_{0.15}}$<br>$\mathbf{-1.19_{0.04}}$ | $\mathbf{-3.62_{0.4}}$<br>$\mathbf{+2.49_{0.2}}$ | $\mathbf{+0.66_{0.1}}$<br>**-** | $\mathbf{+0.53_{0.03}}$<br>$\mathbf{+0.52_{0.03}}$ | $\mathbf{-0.35_{0.03}}$<br>$\mathbf{-0.36_{0.03}}$ | $\mathbf{+0.11_{0.1}}$<br>$\mathbf{+0.11_{0.1}}$ |
| **best**[b] β=2<br>β=0 | $M_{molgas}/M_*$<br>W14 | 1309 | | $\mathbf{+0.16_{0.15}}$<br>$\mathbf{-1.25_{0.03}}$ | $\mathbf{-3.69_{0.4}}$<br>$\mathbf{+2.6_{0.25}}$ | $\mathbf{+0.65_{0.1}}$ | $\mathbf{+0.52_{0..03}}$<br>$\mathbf{+0.53_{0.03}}$ | $\mathbf{-0.36_{0.03}}$<br>$\mathbf{-0.36_{0.03}}$ | **-** |
| all (error w.)<br>β=2 in log(1+z)<br>β=0 | $\mu$=$M_{molgas}/M_*$<br>S14<br>fit *sSFR*<br>instead of $\delta MS$ | 1309 | 0.7<br><br>0.95 | $+0.07_{0.15}$<br><br>$-1.35_{0.03}$ | $-3.8_{0.4}$<br><br>$+2.95_{0.08}$ | $+0.63_{0.1}$<br><br>- | $+0.53_{0.03}$<br><br>$+0.54_{0.03}$ | $-0.33_{0.03}$<br><br>$-0.31_{0.03}$ | - |

Notes to Tables 3a and 3b:

[a] after introduction of zero-points for each method:
zero(CO)= +0.03 dex
zero (FIR Berta/Magnelli)= -0.22
zero(FIR Santini)= +0.21
zero (FIR Bethermin)= -0.023
zero (1mm)= -0.003
[b] zero point offset for $\beta$=2. If the data are fit with $\beta$=0$\rightarrow$ F=0, such that

$$\log M_{molgas}/M_{*} = A + B*\log(1+z) + C*\log(sSFR/sSFR(MS,z,M_{*})) + D*(\log M_{*}-10.7) + E*\log(R_e/<R_e(z,M_{*})>).$$

[c] $R_{e0}$ is the mean effective radius of the star forming population as a function of $z$ and $M_{*}$, as derived by van der Wel (2014) based on the HST CANDELS data: $R_{e0}$=8.9 kpc $(1+z)^{-0.75}(M_{*}/5\times10^{10}M_{\odot})^{0.22}$

**Table A1: parameter distributions extracted from data models**

| Input data set | $z$ | N | $logM_*$ | $log\delta MS$ | depend. variable | $A\text{-}A_{in}$ | $A\text{-}\ B_{in}$ | $C\text{-}C_{in}$ | $D\text{-}D_{in}$ | $G\text{-}G_{in}$ | $H\text{-}H_{in}$ |
|---|---|---|---|---|---|---|---|---|---|---|---|
| $A_{in}$=-1.19,<br>$B_{in}$=2.49,<br>$C_{in}$=0.52,<br>$D_{in}$=-0.36 | $1.2_{0.3}$<br>$2.3_{0.3}$ | 20<br>20 | $10.7_{0.65}$<br>$10.7_{0.65}$ | $0_{0.7}$<br>$0.2_{0.7}$ | log $\mu$ (Figure A1) | $0.02_{0.13}$ | $-0.019_{0.31}$ | $-0.065_{0.12}$ | $0.093_{0.17}$ | - | - |
| $A_{in}$-$D_{in}$ as above | $1.2_{0.3}$<br>$2.3_{0.3}$ | 20<br>20 | $10.7_{0.4}$<br>$10.7_{0.4}$ | $0.4_{0.4}$<br>$0.4_{0.4}$ | log $\mu$ | $0.45_{0.18}$ | $-0.48_{0.29}$ | $-0.35_{0.12}$ | $0.085_{0.2}$ | - | - |
| $A_{in}$=-1.42,<br>$B_{in}$=3.54,<br>$C_{in}$=0.52,<br>$D_{in}$=-0.36,<br>$G_{in}$=-3.25 | $0.02_{0.01}$<br>$0.02_{0.01}$<br>$0.2_{0.1}$<br>$0.6_{0.2}$<br>$1.3_{0.4}$<br>$2.3_{0.3}$<br>$3.4_{0.5}$ | 200<br>100<br>20<br>120<br>100<br>80<br>20 | $10.7_{0.65}$<br>$10.7_{0.4}$<br>$10.7_{0.4}$<br>$10.7_{0.65}$<br>$10.7_{0.65}$<br>$10.7_{0.65}$<br>$10.7_{0.4}$ | $0_{0.8}$<br>$0.5_{0.3}$<br>$0.5_{0.5}$<br>$0_{0.8}$<br>$0.2_{0.8}$<br>$0.4_{0.8}$<br>$0.4_{0.4}$ | log $\mu$ | $0.001_{0.013}$ | $-0.002_{0.14}$ | $-0.035_{0.03}$ | $0.027_{0.02}$ | $0.43_{1.6}$ | - |
| $A_{in}$= -1.42,<br>$B_{in}$= 3.54,<br>$C_{in}$= 0.52,<br>$D_{in}$= -0.36,<br>$G_{in}$= -3.25,<br>$H_{in}$= -0.1 | $0.02_{0.01}$<br>$0.02_{0.01}$<br>$0.2_{0.1}$<br>$0.6_{0.2}$<br>$1.3_{0.4}$<br>$2.3_{0.3}$<br>$3.4_{0.5}$ | 200<br>100<br>20<br>120<br>100<br>80<br>20 | $10.7_{0.65}$<br>$10.7_{0.4}$<br>$10.7_{0.4}$<br>$10.7_{0.65}$<br>$10.7_{0.65}$<br>$10.7_{0.65}$<br>$10.7_{0.4}$ | $0_{0.8}$<br>$0.5_{0.3}$<br>$0.5_{0.5}$<br>$0_{0.8}$<br>$0.2_{0.8}$<br>$0.4_{0.8}$<br>$0.4_{0.4}$ | log $\mu$ | $-0.001_{0.01}$ | $-0.03_{0.12}$ | $-0.03_{0.013}$ | $0.01_{0.01}$ | $0.09_{0.23}$ | $0.01_{0.01}$ |
| $A_{in}$=+0.086,<br>$B_{in}$=-0.6,<br>$C_{in}$=-0.44,<br>$D_{in}$=+0.056 | $1.2_{0.3}$<br>$2.3_{0.3}$ | 20<br>20 | $10.7_{0.65}$<br>$10.7_{0.65}$ | $0.2_{0.8}$<br>$0.3_{0.8}$ | log $t_{depl}$ | $-0.02_{0.08}$ | $0.04_{0.2}$ | $-0.029_{0.05}$ | $-0.013_{0.05}$ | - | |
| $A_{in}$=+0.086,<br>$B_{in}$=-0.6,<br>$C_{in}$=-0.44,<br>$D_{in}$=+0.056 | $2.3_{0.3}$<br>$3.3_{0.3}$ | 30<br>10 | $10.7_{0.5}$<br>$10.7_{0.5}$ | $0.5_{0.4}$<br>$0.5_{0.4}$ | log $t_{depl}$ | $-0.02_{0.24}$ | $0.04_{0.45}$ | $-0.029_{0.05}$ | $-0.013_{0.09}$ | | |
| $A_{in}$= -1.42,<br>$B_{in}$= 3.54,<br>$C_{in}$= 0.52,<br>$D_{in}$= -0.36,<br>$G_{in}$= -3.25,<br>$H_{in}$= -0.1 | $0.22_{0.1}$<br>$0.5_{0.3}$ | 20<br>20 | $10.7_{0.65}$<br>$10.7_{0.65}$ | $0.0_{0.6}$<br>$0.0_{0.6}$ | log $\mu$ | $-1.407_{0.08}$ | $3.65_{0.75}$ | $0.48_{0.06}$ | $-0.32_{0.06}$ | | |
| $A_{in}$-$H_{in}$ as above | $1.2_{0.3}$<br>$2.3_{0.3}$ | 20<br>20 | $10.7_{0.65}$<br>$10.7_{0.65}$ | $0.0_{0.6}$<br>$0.0_{0.6}$ | log $\mu$ | $-0.89_{0.14}$ | $1.81_{0.32}$ | $0.46_{0.06}$ | $-0.28_{0.06}$ | | |
| $A_{in}$-$H_{in}$ as above | $2.3_{0.3}$<br>$3.3_{0.3}$ | 20<br>20 | $10.7_{0.65}$<br>$10.7_{0.65}$ | $0.0_{0.6}$<br>$0.0_{0.6}$ | log $\mu$ | $-0.43_{0.17}$ | $0.94_{0.31}$ | $0.47_{0.06}$ | $-0.29_{0.06}$ | | |